\pdfoutput=1

\documentclass[11pt]{article}

\usepackage[preprint]{acl}

\usepackage{times}
\usepackage{latexsym}
\usepackage{float}
\usepackage{paracol} 
\usepackage{annotation}

\usepackage{booktabs}
\usepackage{tablefootnote}
\usepackage{ai-usage-card}
\usepackage{cleveref}
\usepackage[T1]{fontenc}

\usepackage[utf8]{inputenc}

\usepackage{microtype}
\usepackage{inconsolata}

\usepackage{graphicx}
\usepackage{multicol}

\usepackage[nolist]{acronym}
\usepackage{caption}
\usepackage{subcaption}

\usepackage{colortbl}
\usepackage{xcolor} %
    \definecolor{UBlau}{HTML}{153268}
    \definecolor{LBlau}{HTML}{005f9b}
    \definecolor{LMBlau}{HTML}{0091c8}
    \definecolor{LHBlau}{HTML}{50a5d2}
    \definecolor{WiWi}{HTML}{2b7ab3}
    \definecolor{Grau60}{HTML}{878786}
\usepackage{pgfplots}
\usepackage{collcell}

\newcommand{\gradientcell}[3]{%
  \pgfmathsetmacro{\percent}{min(max((#1-#2)/(#3-#2)*100,0),100)}%
  \edef\tempcellcolor{\noexpand\cellcolor{violet!\percent!lightgray!50!}}%
  \tempcellcolor #1%
}

\usepackage{tcolorbox}
\newtcolorbox{combinedprompt}{
    colframe=black,
    colback=white,
    boxrule=0.6mm,
    width=\linewidth,
    fonttitle=\bfseries,
    rounded corners,
    coltitle=black
}

\usepackage{listings} %
\lstnewenvironment{configpython}[2][] %
{
    \lstset{
        commentstyle=\color{UBlau},
        keywordstyle=\color{WiWi},
        numberstyle=\tiny\color{Grau60},
        basicstyle=\ttfamily\footnotesize,
        language=python,
        numbers=none,
        frame=tblr,           %
        tabsize=4,
        captionpos=b,         %
        caption={#1},
        label={#2},
        breaklines=true,      %
        breakatwhitespace=false,
        breakautoindent=true,
        prebreak=\mbox{\textcolor{red}{$\hookrightarrow$}} %
    }
}{}

\lstnewenvironment{configjson}[2][] %
{
    \lstset{
        commentstyle=\color{UBlau},
        keywordstyle=\color{WiWi},
        numberstyle=\tiny\color{Grau60},
        basicstyle=\ttfamily\footnotesize,
        numbers=none,
        frame=tblr,           %
        tabsize=4,
        captionpos=b,         %
        caption={#1},
        label={#2},
        breaklines=true,      %
        breakatwhitespace=false,
        breakautoindent=true,
        prebreak=\mbox{\textcolor{red}{$\hookrightarrow$}} %
    }
}{}

\definecolor{systemcolor}{rgb}{0.9,0.9,1} %
\definecolor{usercolor}{rgb}{1,0.9,0.9} %

\title{Voting or Consensus? Decision-Making in Multi-Agent Debate}

\author{
 \textbf{Lars Benedikt Kaesberg\textsuperscript{1,*}},
 \textbf{Jonas Becker\textsuperscript{1,2}},
 \textbf{Jan Philip Wahle\textsuperscript{1}},
 \textbf{Terry Ruas\textsuperscript{1}},
 \textbf{Bela Gipp\textsuperscript{1}}
\\[3pt]
 \textsuperscript{1}University of Göttingen, Germany; \textsuperscript{2}LKA NRW, Germany
\\
 \small{
   \textbf{\textsuperscript{*}Correspondence:} \href{mailto:l.kaesberg@uni-goettingen.de}{l.kaesberg@uni-goettingen.de}
 }
}
\begin{acronym}
  \acro{AAD}{All-Agents Drafting}
  \acro{CI}{Collective Improvement}
  \acro{CoT}{Chain-of-Thought}
  \acro{fpc}{finite population correction}
  \acro{LLM}{large language model}
  \acro{MALLM}{\textbf{M}ulti-\textbf{A}gent \textbf{L}arge \textbf{L}anguage \textbf{M}odels}
\end{acronym}
\begin{document}
\maketitle
\AddAnnotationRef

\begin{abstract}

Much of the success of multi-agent debates depends on carefully choosing the right parameters.
The decision-making protocol stands out as it can highly impact final model answers, depending on how decisions are reached. 
Systematic comparison of decision protocols is difficult because many studies alter multiple discussion parameters beyond the protocol.
So far, it has been largely unknown how decision-making influences different tasks.
This work systematically evaluates the impact of seven decision protocols (e.g., majority voting, unanimity consensus).
We change only one variable at a time --- the decision protocol --- to analyze how different methods affect the collaboration between agents and measure differences in knowledge and reasoning tasks.
Our results show that voting protocols improve performance by $13.2\%$ in reasoning tasks and consensus protocols by $2.8\%$ in knowledge tasks compared to other decision protocols.
Increasing the number of agents improves performance, while more discussion rounds before voting reduce it.
To improve decision-making by increasing answer diversity, we propose two new methods, \ac{AAD} and \ac{CI}.
Our methods improve task performance by up to $3.3\%$ with \ac{AAD} and up to $7.4\%$ with \ac{CI}. 
This work demonstrates the importance of decision-making in multi-agent debates beyond scaling.

\end{abstract}

\section{Introduction}
\label{sec:intro}

\begin{figure}[!ht]
    \centering
    \includegraphics[width=\columnwidth]{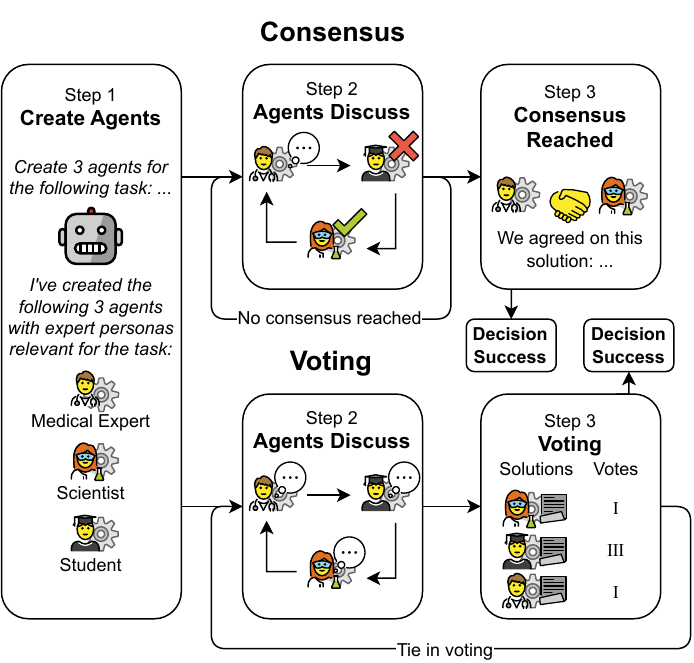}
    \caption{Illustration of voting and consensus-based decision protocols used in this study.
    \vspace{-0.3cm}
    }
    \label{fig:decision_protocols}
\end{figure}

Humans are inherently social, and collaboration has been key to innovation and progress. %
We know that generating solutions together is only beneficial if we can effectively select, agree, and commit to them.
History, sociology, and psychology have long demonstrated how different decision-making processes influence collective outcomes \citep{jones_comparison_1994, list_social_2022}. 
Multi-agent systems form a parallel to human behavior by solving problems collectively via debate.
However, so far, few studies have investigated how decision-making influences \ac{LLM} collaboration and their ability to problem-solve.

Current approaches often apply decision strategies like majority voting \citep{yang_llm_2024} or consensus \citep{yin_exchange--thought_2023} indiscriminately to various tasks.
Consensus strategies refine decisions while voting approaches choose from proposed solutions.
Prior work often treats these strategies as fixed variables without considering problem-specific characteristics \citep{yin_exchange--thought_2023}.
We show that changes in decision protocols lead to markedly different results among tasks.
Thus, we argue that decision-making is central to multi-agent processes.

This study systematically quantifies the effectiveness of decision-making protocols within multi-agent debates on knowledge tasks (i.e., MMLU \citep{hendrycks_measuring_2021-1}, MLLU-Pro \citep{wang_mmlu-pro_2024}, GPQA \citep{rein_gpqa_2023}) and reasoning tasks (i.e., SQuAD 2.0 \citep{rajpurkar_know_2018}, StrategyQA \citep{geva_did_2021}, MuSR \citep{sprague_musr_2024}). 
In contrast to prior work, which varies the decision protocol concurrently to other parameters~\citep{yang_llm_2024, yin_exchange--thought_2023}, we use a consistent setup to quantify changes for which only the decision mechanism is responsible.
Specifically, we explore the impact of three consensus \citep{chen_reconcile_2024} and four voting \citep{yang_llm_2024} methods and highlight their task-specific strengths and weaknesses.
An overview of how consensus and voting decision protocols work in our setup can be seen in \Cref{fig:decision_protocols}.
Throughout our experiments, we observe that answer diversity and independent answer generation have a marked impact on decision-making. 
To reap the benefits of these factors, we propose two methods: \acf{AAD} and \acf{CI}.
\ac{AAD} ensures each agent independently drafts an initial solution before any interaction, promoting distinct reasoning paths. %
\ac{CI} structures agent collaboration through iterative refinement while preventing excessive communication to avoid bias towards similar answers between agents.

Our experiments show that consensus protocols perform better in knowledge tasks, improving performance by $2.8\%$, and voting protocols perform better in reasoning tasks, improving results by $13.2\%$. 
\ac{AAD} increases accuracy by about $3.3\%$, while \ac{CI} further improves effectiveness, leading to an $7.4\%$ performance boost. 
Our results underline the role of decision protocols for multi-agent experiments. 
We recommend consensus strategies for knowledge tasks and voting for reasoning tasks while generally using \ac{AAD} or \ac{CI} to improve answer diversity. 
Our investigations into the role of decision protocols are particularly relevant for high-stakes domains such as medical diagnostics and legal reasoning, where wrong decisions can have real-world negative effects. 
We release the code and data for these experiments publicly\footnote{\href{https://github.com/lkaesberg/decision-protocols}{github.com/lkaesberg/decision-protocols}}. 

\medskip
\noindent\textbf{Key Contributions:}
\begin{itemize}
    \item[§\ref{sec:experiment1}] A systematic comparison of decision protocols, revealing for the first time that consensus is most effective for knowledge tasks, while voting is better in reasoning tasks.
    \item[§\ref{sec:experiment-numagents}] An analysis of scaling differences between increasing the number of agents and extending the number of communication rounds.
    \item[§\ref{sec:experiment2}] Two new methods,\ac{AAD} and \ac{CI}, to enhance answer diversity in multi-agent debates by encouraging independent thinking.
\end{itemize}

\section{Related Work}

\noindent
\textbf{LLMs as Agents}. %
An agent differs from an LLM in that it has a defined planning behavior, can use tools, and maintains a state or memory across interactions. 
Techniques such as \acf{CoT} prompting \citep{wei_chain--thought_2023}, self-refinement \citep{madaan_self-refine_2023}, and self-consistency \citep{wang_self-consistency_2022} improve models' ability to plan, critique, and refine responses. 
Persona-based prompting \citep{jiang_personallm_2024} enables \acp{LLM} to adopt specialized roles, improving answer diversity.
A single-agent system operates as one entity with an internal state, while a multi-agent debate consists of multiple agents with a private state that persists across calls and may operate asynchronously, leading to emergent, independent behaviors \cite{du2023improvingfactualityreasoninglanguage, ZhaoHXL23a, XuYLW23a, SuzgunK24a, goldberg2024}.
In multi-agent debates, many parameter choices have to be made, such as in which turn order they communicate \citep{yin_exchange--thought_2023}, and which tools they can use \citep{yao2023reactsynergizingreasoningacting}.
Yet one choice is inevitable: How to make decisions between agents.

\noindent
\textbf{Decision-Making}. Finding collective solutions markedly impacts human decision-making \cite{jones_comparison_1994}. 
While consensus promotes shared decisions and allows everyone to contribute to the final solution, it can be time-consuming and lead to power concentration of individuals with ``vetos''. 
Conversely, voting can lead to a faster final decision because it streamlines decision-making, but is susceptible to manipulation and does not take into account all opinions.
For example, participants can vote for a less preferred but more viable alternative to block an undesired outcome, leading to outcomes that do not reflect the collective will of the group \citep{list_social_2022}.

Research in multi-agent debate has implemented various human decision protocols \citep{yin_exchange--thought_2023, chen_reconcile_2024, yang_llm_2024}. 
Exchange-of-Thought \citep{yin_exchange--thought_2023} employs a consensus-based approach, where agents iteratively refine answers through discussion in reasoning tasks, but they do not compare consensus to other decision protocols.
\citet{yang_llm_2024, 10.1145/3665332} introduce multiple voting protocols (e.g., approval voting) but do not compare the performance of these voting decision protocols across different tasks, focussing more on a comparison of how humans vote compared to \acp{LLM}.
ReConcile \citep{chen_reconcile_2024} integrates a hybrid voting and consensus approach by iteratively refining answers through confidence-weighting until consensus is reached. %
In summary, prior work focuses on a single class of decision protocols without systematic comparison, partly because many parameters change between experiments \citep{yin_exchange--thought_2023, yang_llm_2024, chen_reconcile_2024}.
This work systematically evaluates seven voting and consensus approaches on both knowledge and reasoning datasets to show task-based advantages of one protocol over another.

\section{Methodology}

In the following, we explain the multi-agent environment, decision protocols, response generators, and datasets used in our experiments. 

\subsection{Setup}
\label{sec:setup}
We run multi-agent debates based on three key components: a discussion paradigm, a decision protocol, and an agent response generator.
Each discussion consists of three automatically generated expert personas for the task following \citet{kim_persona_2024}. 
While the number of personas can vary, \citet{yin_exchange--thought_2023} found this number to be the most efficient.
Using personas is important as it generates agents with expertise in different domains, incorporating diverse viewpoints for the final decision.
After that, the agents discuss the problem for multiple turns.
The number of turns varies from one to five, depending on the experiment setup and decision protocol.
The \textit{decision protocol} defines what criteria must be met for the discussion to end and how the final solution will be created.
Each agent can generate one answer per turn and, based on the agent's turn order, respond to the other agents as a default behavior.
If not explicitly stated, all agents exchange messages with one another.
These discussion characteristics are defined by the \textit{discussion paradigm}.
For consensus decision protocols, each agent also indicates whether they agree with the previous message.
We repeat this process each round until a certain level of agreement (defined by the decision protocol) is reached, which leads to the final solution. 
For voting decision protocols, the agents start voting after the third turn.
If they successfully agree on a solution in any turn after the third, the discussion ends.
Agents can only memorize messages from up to two turns to limit the context length provided to the agent.
We use a \textit{response generator} to define how agents respond to previous messages.
It determines how the discussion history is presented, what additional information is provided, how the persona is introduced, and in which tone they should respond (e.g., neutral or critical).
Additional details on the experimental setup can be found in \Cref{sec:mallm}.

The \acp{LLM} used for the experiments are Llama 3 8B and 70B\footnote{The two models used for this study can be found at \href{https://huggingface.co/meta-llama/Meta-Llama-3-8B-Instruct}{meta-llama/Meta-Llama-3-8B-Instruct} and \href{https://huggingface.co/meta-llama/Meta-Llama-3-70B-Instruct}{meta-llama/Meta-Llama-3-70B-Instruct}} \citep{dubey_llama_2024}.
Within one discussion, all prompted agents use the same base model. 
The smaller model uses two NVIDIA RTX5000 with 16 GB of VRAM (for $\sim250$ hours), and the larger model uses eight NVIDIA A100 with 40GB VRAM (for $\sim40$ hours). 
A list of experiment-specific parameters is available in \Cref{sec:mallm_setup_app}.

\subsection{Agent Prompts and Decision Protocols}
\label{sec:decision_protocols_main}

\begin{figure*}
    \centering
    \includegraphics[width=.85\linewidth]{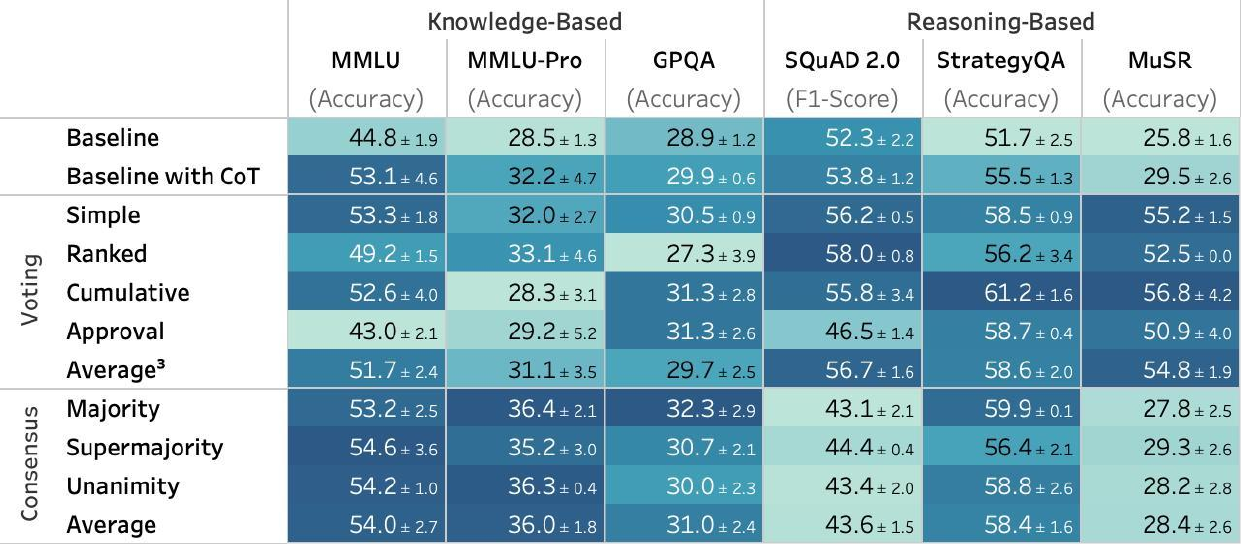}
    \caption{Task performance{\tiny$\pm$std} for seven decision protocols (voting and consensus-based) on six tasks (knowledge and reasoning) based on agents with Llama 8B. \textbf{Bold} indicates the highest results per dataset. Standard deviation for three runs.}
    \label{tab:results_8b}
\end{figure*}

Prompt design has a marked impact on multi-agent debate.
For this work, we propose three response generators.
The \textbf{Simple Response Generator} is our default in which agents are prompted to answer neutral and unbiased to previous messages. 
The \textbf{Critical Response Generator} encourages agents to critically assess prior answers and propose new solutions, countering sycophantic tendencies \cite{sharma_towards_2023}. 
The \textbf{Reasoning Response Generator} restricts agents to sharing only reasoning paths, thus avoiding bias towards agreeing with other agents' final solutions.

Decision protocols then determine the final solution of the discussion by selecting the most promising solution based on predefined mechanisms. 
We use two classes of protocols in this work. 

\textbf{Consensus decision protocols} decide on an answer by prompting the agents to converge on one shared solution. 
The solution is selected when a required level of agreement among agents is reached.
We explore three major agreement levels: \textit{majority consensus} - more than $50\%$ agreement, \textit{supermajority consensus} - more than $66\%$ agreement, and \textit{unanimity consensus} - all agents have to agree.
We extend the majority consensus of \citet{yin_exchange--thought_2023} by higher agreement levels, i.e., supermajority (66\%), unanimity (100\%).

\textbf{Voting decision protocols} allow several possible solutions to be presented in parallel during the discussion, and ultimately all agents must vote on a final solution.
If there is a tie in the voting, all agents will discuss for another round and then vote again.
Our work includes four different voting decision protocols inspired by \citet{yang_llm_2024}:
\textit{simple voting} - each agent casts one vote, and the answer with the most votes wins;
\textit{ranked voting} - each agent ranks the solutions from best to worst to find the solution that consistently has the highest rank over all individual rankings;
\textit{approval voting} - each agent casts unlimited votes, and the answer with the most votes wins; and
\textit{cumulative voting} - each agent is allowed to divide up to 25 points between all solutions. The solution with the most points wins.

\subsection{Datasets}
\label{sec:datasets_main}

We evaluate our decision protocols using six datasets: three knowledge tasks (i.e., MMLU, a broad-topic test; MMLU-Pro, a domain-specific test with challenging questions; GPQA, a specialized question set difficult for web search) and three reasoning tasks (i.e., StrategyQA, a multistep reasoning task; MuSR, a long-context murder mystery; SQuAD 2.0, a reading comprehension task with questions that may or may not have answers in the context).
Because of computational constraints, we calculate the task performance on a subset of samples with three runs and report their standard deviation.
More details about the datasets, the number of samples, and the sampling strategy can be found in \Cref{appendix:datasets}.

\section{Experiments}
We use a set of diverse reasoning and knowledge tasks to explore the effectiveness and limits of decision protocols in multi-agent debate.

\stepcounter{footnote}\footnotetext{Approval Voting is left out as it consistently fails to reach a voting decision as described in \Cref{sec:experiment1}.}

\subsection{Performance of Decision Protocols}
\label{sec:experiment1}

Multi-agent debates for task solving have shown promise in recent research, but they rely on a varying choice of decision protocols, hindering systematic comparison \citep{yin_exchange--thought_2023, chen_reconcile_2024}. 
This experiment systematically compares a set of seven decision protocols, changing only the decision protocol used and nothing else.
Specifically, we determine whether some protocols have advantages over others and how they behave in specific edge cases, such as for unanswerable questions.

We experiment with four voting-based (Simple, Ranked, Cumulative, Approval) and three consensus-based (Majority, Supermajority, Unanimity) methods and test them on three knowledge tasks (MMLU, MMLU-Pro, GPQA) and three reasoning tasks (SQuAD 2.0, StrategyQA, MuSR).
We use the Llama 3 8B model with and without \ac{CoT} prompting to generate baseline results for comparison with multi-agent debate.
We compare answerable and unanswerable questions of the SQuAD 2.0 dataset to inspect how decision protocols behave in edge cases.

\begin{figure}[t]
    \centering
    \includegraphics[width=\linewidth]{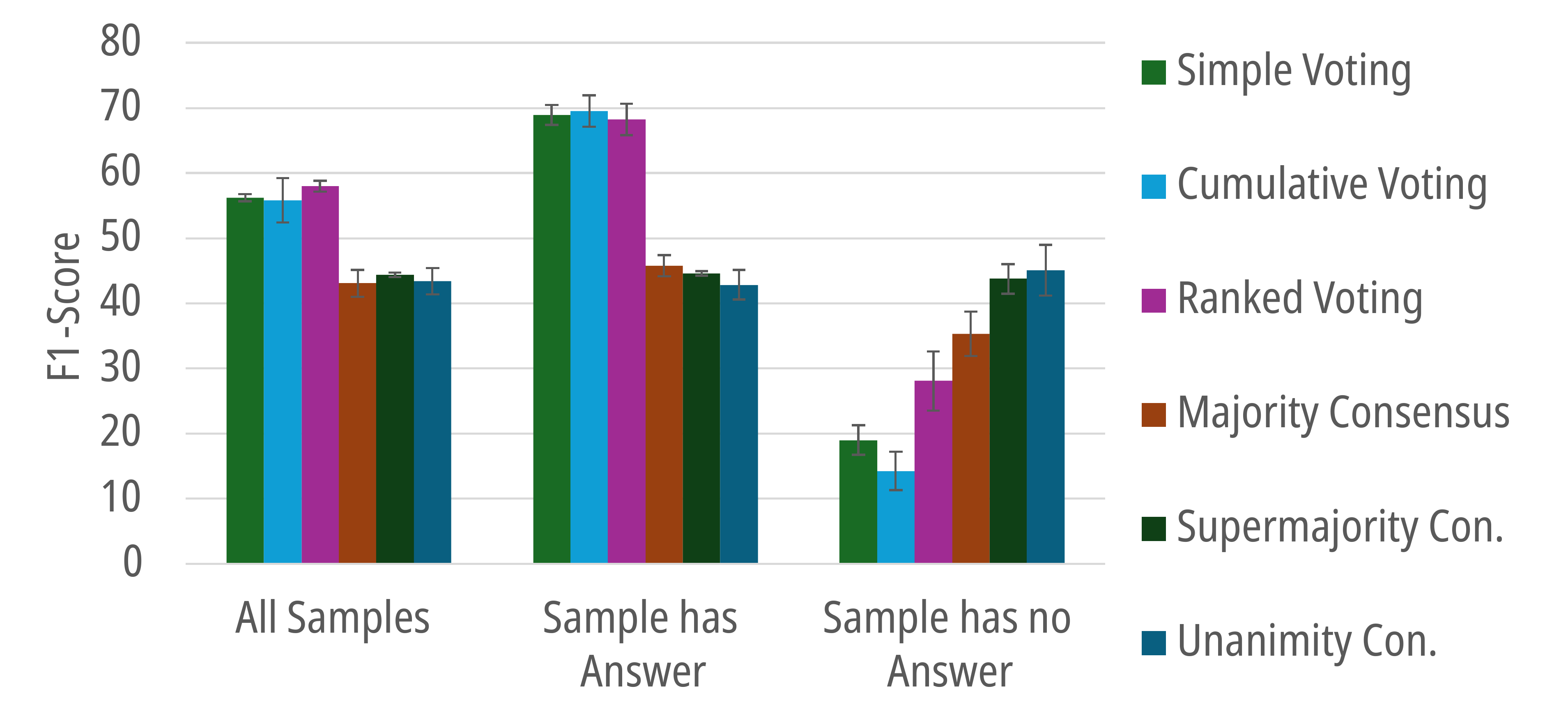}
    \caption{F1{\tiny$\pm$std} of different decision protocols on SQuAD 2.0 divided into three ablation groups: (middle) samples with an answer in the context (\textbf{Sample has Answer}), (right) samples with no answer in the context (\textbf{Sample has no Answer}), and (left) the combination of both (\textbf{All Samples}). Standard deviation over three runs.
    }
    \label{fig:squad_divided}
\end{figure}

\Cref{tab:results_8b} shows the baselines compared to multi-agent debate, with the decision protocols being the rows grouped by voting and consensus-based protocols.
The columns are the datasets grouped by knowledge and reasoning-based tasks.
Overall, multi-agent debate with consensus and voting decision protocols outperforms the \ac{CoT} baseline. 
Notably, consensus protocols perform better on knowledge-based tasks with average improvements of $2.3\%$ on MMLU, $4.9\%$ on MMLU-Pro, and $1.3\%$ on GPQA over the voting average.
Voting protocols perform better in reasoning tasks with average improvements of $13.1\%$ on SQuAD 2.0, $0.2\%$ on StrategyQA, and $26.4\%$ on MuSR over the consensus average.
The results for the Llama 3 70B model can be found in \Cref{app:decision_70b}.
On average, voting-based protocols require $3.38$ rounds and consensus-based protocols $1.42$ rounds to reach a decision, indicating that consensus is more suitable for applications that require quick decision-making as they use less test-time compute.
Interestingly, most conversations reach a decision before they are terminated after five turns.
Approval voting is the only protocol that often leads to no agreed solution ($59\%$ of cases) due to the agreeableness of the agents. 
The agents often simply vote for all answers as they do not have to decide, leading to many ties (see \Cref{app:termination_percent} for details). 
\Cref{fig:squad_divided} shows the F1-Score for SQuAD 2.0 divided into whether questions are answerable or not for each decision protocol.
Voting achieves a higher overall F1-score (56.7\%) than consensus (43.6\%), especially for answerable samples.
However, consensus is more effective for unanswerable samples, and stricter consensus methods (e.g., unanimity consensus) produce even higher scores.

The effectiveness of decision protocols depends on the nature of the task. 
Our results suggest that consensus excels in knowledge tasks, and voting works better for reasoning tasks.
Voting protocols might outperform consensus in reasoning tasks because they allow the exploration of multiple reasoning paths. %
One reason for the good performance of consensus in knowledge tasks is that it mitigates individual agent errors by requiring multiple agents to agree on the same statement before reaching a conclusion. 
This approach is less susceptible to manipulation or errors by any single agent. 
We found voting can be easily tricked by generating answers that seem to be correct but have some incorrect statements (see \Cref{example:trick_discussion}).
The consensus method has a distinct advantage by incorporating a safeguard of repeated checks across agents to find these small errors.
The detailed analysis of SQuAD 2.0 further supports the finding that decision protocol effectiveness is task-dependent.

Both experiments reveal that the decision protocol used can significantly impact the overall performance of multi-agent debates.
While multi-agent approaches generally require more computation than a single agent (approximately five times the compute resources for consensus protocols and ten times for voting protocols compared to the CoT baseline in our setup), the choice between protocols offers distinct advantages beyond simply scaling resources.
Notably, consensus protocols often reached a decision faster than voting protocols, particularly on knowledge tasks that also achieved superior performance (\Cref{tab:results_8b}).
Conversely, voting excelled on reasoning tasks despite potentially needing more rounds and compute. 
This demonstrates that with the correct choice, multi-agent debates can be made more error-resistant by using consensus decision protocols and more explorative by using voting decision protocols. 
However, we acknowledge that these performance improvements must be weighed against the substantial increase in computational resources required compared to simpler baseline methods. 
High token consumption is a significant practical limitation of the multi-agent debate (MAD) paradigm, and its prevalence in the field does not diminish its importance as a challenge. 
Observed performance improvements, although positive, may not always scale proportionally with the increased computational resources.
Our study's primary objective was to systematically compare how different decision protocols function within this resource-intensive MAD paradigm to help researchers select the most effective protocol for their specific tasks.  
Therefore, researchers should carefully consider whether the incremental performance gains justify the associated computational costs in their particular use cases.
An additional experiment comparing solution counting (selecting the answer that is given the most) to our prompted decision protocols showed similar trends and had little impact on our main findings (see \Cref{sec:answer-counting-ablation}).

\subsection{Number of Agents and Discussion Rounds}
\label{sec:experiment-numagents}
Research by \citet{yin_exchange--thought_2023} suggests that increasing the number of rounds and agents participating in a multi-agent debate is beneficial because of test-time compute scaling.
In our experimental setup, agents communicate for three rounds before voting, following \citet{du_improving_2023}. 
As our first experiment showed, consensus reaches a decision much faster than voting.
Here, we investigate how the number of discussion rounds before voting impacts task performance, as fewer rounds with the same accuracy would improve computational efficiency.
\citet{becker_multi-agent_2024} suggests that extended discussions may lead to decreased accuracy because of agents drifting away from the original task. 
\citet{wang_rethinking_2024} shows increasing the number of agents may have positive effects. 

We conduct this experiment using the StrategyQA dataset because multi-agent debate has consistently mitigated errors of single agents using \ac{CoT} as shown in the previous experiment.
The experiment is structured in two parts.
First, we fix the number of agents to three and increase the number of discussion rounds from one to ten. 
Second, we fix the number of rounds to three and increase the number of agents from one to ten. 
Both experiments use the \textit{simple voting} decision protocol, and each condition is evaluated across three independent runs. 

\begin{figure}[t]
  \centering
  \includegraphics[width=\linewidth]{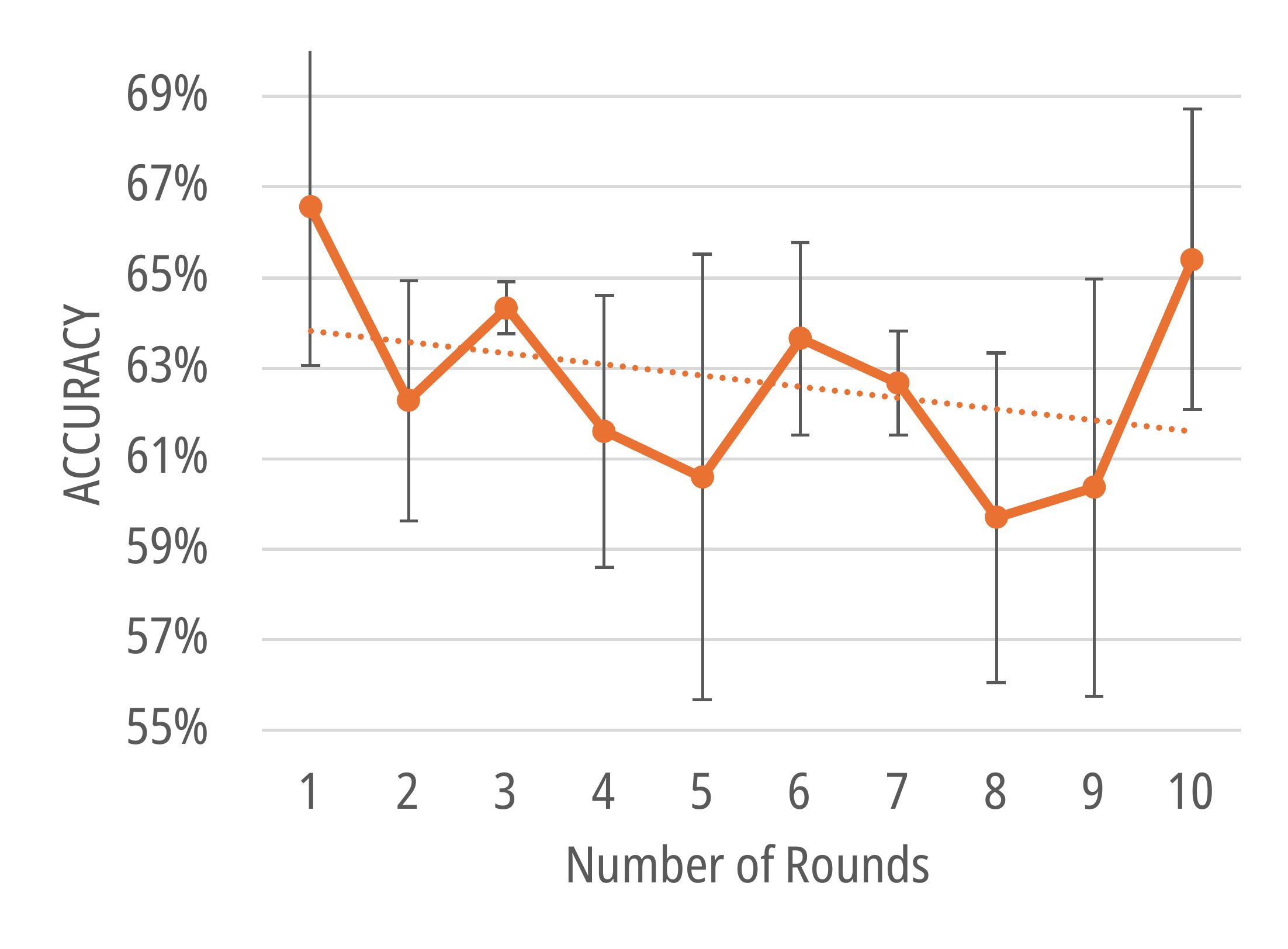}
  \caption{Accuracy{\tiny$\pm$std} on StrategyQA when the agents have to talk for a given number of rounds before they are allowed to vote using the simple voting decision protocol. Standard deviation over three runs.}
  \label{fig:num_rounds}
\end{figure}

\begin{figure}[t]
  \centering
  \includegraphics[width=\linewidth]{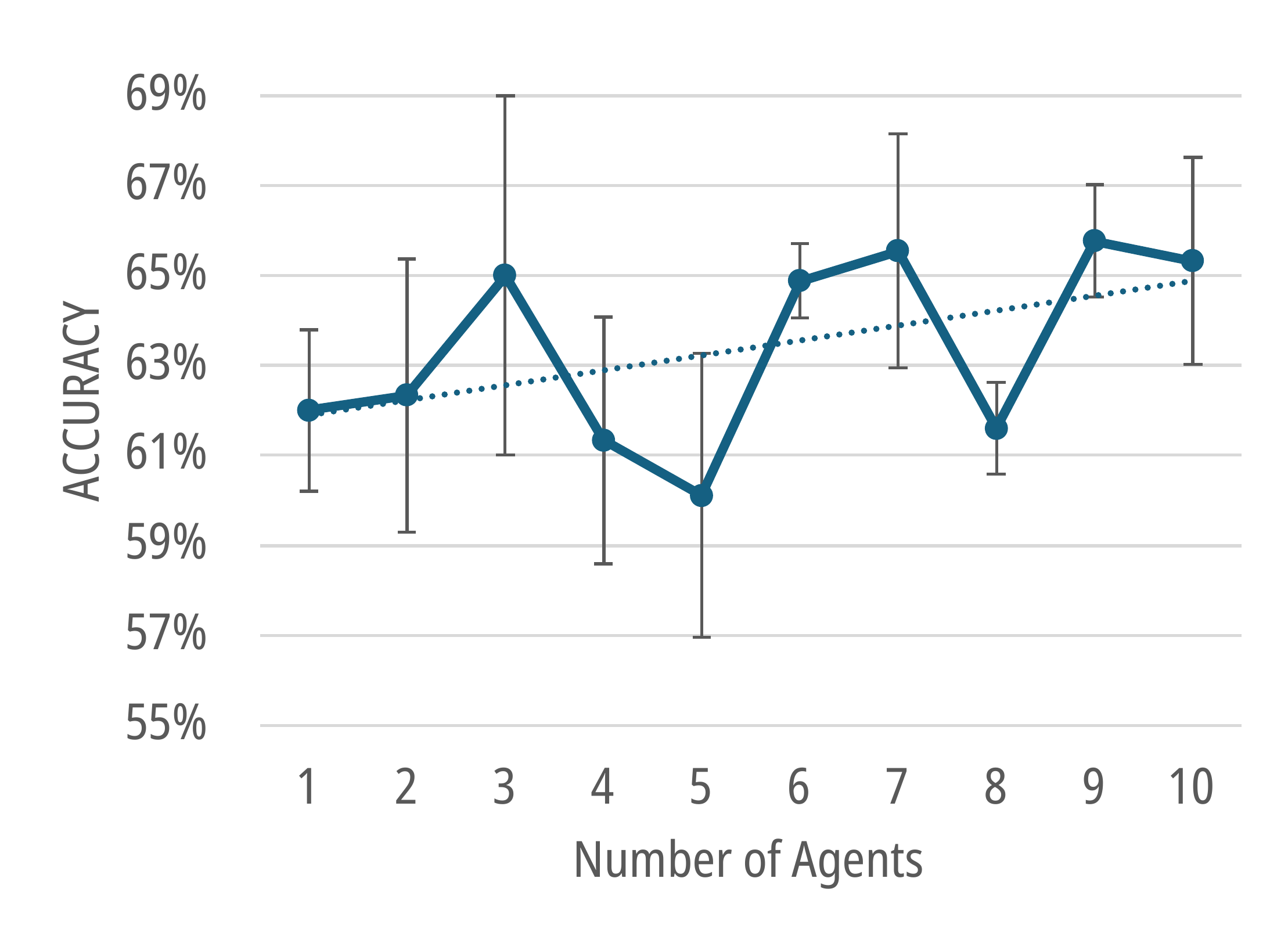}
  \caption{Accuracy{\tiny$\pm$std} on StrategyQA with a different number of agents participating in the discussion. The final answer is created using the simple voting decision protocol. Standard deviation over three runs.}
  \label{fig:num_agents}
\end{figure}

\Cref{fig:num_rounds} shows accuracy relative to the number of rounds the agents have to communicate before voting starts. 
A linear regression, visualized by the dotted line, indicates that increasing discussion rounds slightly reduces accuracy. 
Still, with the increase from nine to ten rounds a rather abrupt increase in task performance can be seen.  
Separately, \Cref{fig:num_agents} shows the task performance relative to the number of agents participating in the debate. 
Here, a linear regression indicates that increasing the number of agents leads to a slight upward trend in accuracy, suggesting that larger groups improve task performance.

Having more agents generating solutions generally leads to higher accuracy.
This resembles the effect seen in self-consistency \citep{wang_self-consistency_2022}, where increasing the number of sampled answers improves task performance.
However, \Cref{fig:num_rounds} shows that increasing the number of debate rounds decreases performance, contradicting the expected benefits of self-refinement.
This aligns with recent concerns about the reliability of self-refinement, as \citet{huang_large_2024} questions the findings of \citet{madaan_self-refine_2023}, suggesting that the claimed improvements may not be robust.

At the same time, we also have one case where we see an increase in task performance as the number of rounds increases. Observe the sharp upside arc in \Cref{fig:num_rounds} for turn ten.
To test whether individual rounds can recover this performance decrease, we create an ablation experiment where agents challenge the final result and propose a new solution.
This extra round allows us to verify the effects of an extended discussion and also if agents would change their final solution.
We create five different challenge scenarios and test them on two knowledge tasks (MMLU and MMLU-Pro) and two reasoning tasks (StrategyQA and SQuAD 2.0).
In the first setting, we provide agents with the final solution and ask whether they agree or want to change it. 
In the second setting, agents can access the full discussion history with prior reasoning. 
The third setting includes additional context we retrieve with RAG from Wikipedia\footnote{We use: \href{https://github.com/Multi-Agent-LLMs/context-plus}{github.com/Multi-Agent-LLMs/context-plus}}.
The fourth and fifth settings act as a baseline, introducing intentionally incorrect solutions---either irrelevant or wrong answers generated by the Qwen2 7B \citep{yang_qwen2_2024} model to test whether agents can detect incorrect solutions. 
Solutions are generated with a different model than Llama 3, to remove any bias that might occur if the same model used to detect the wrong answer also generated it \cite{wahle_are_2021}.

\begin{figure}[t]
    \centering
    \includegraphics[width=\linewidth]{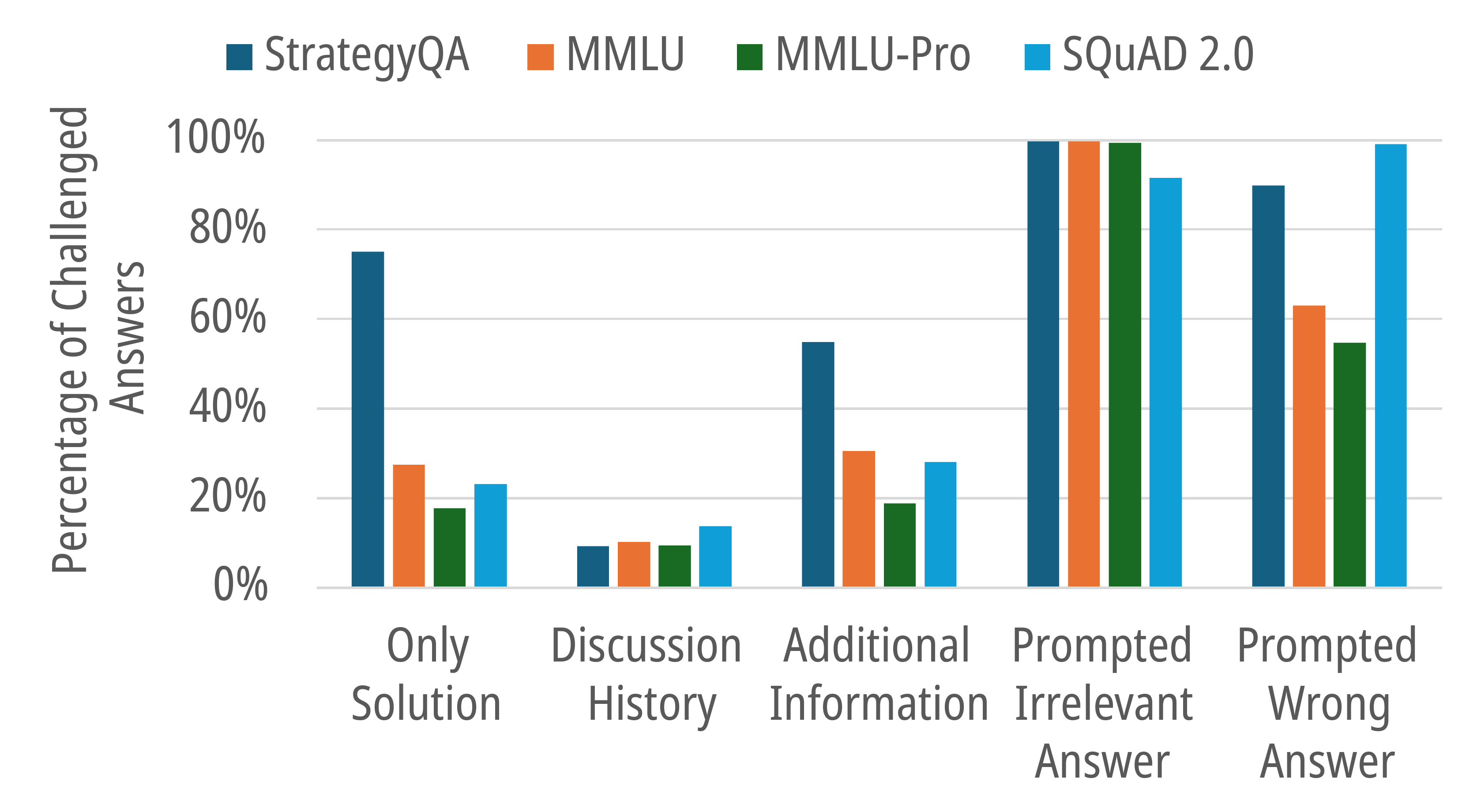}
    \caption{Percentage of agents challenging the final solution with different levels of information. %
   \vspace{-0.3cm}
  }
    \label{fig:challenge}
\end{figure}

\begin{figure}[t]
    \centering
    \includegraphics[width=\linewidth]{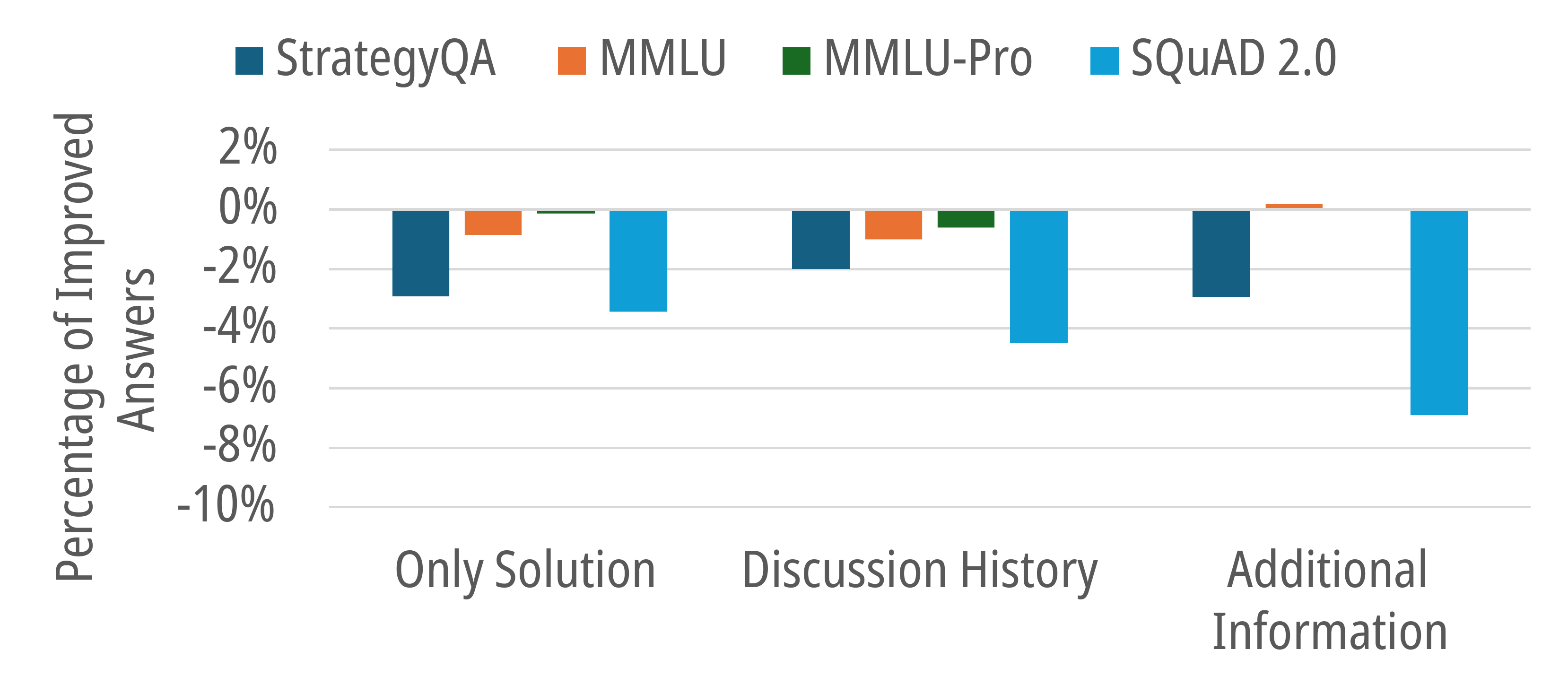}
    \caption{Percentage of answers that improve after being challenged using different levels of information. %
    \vspace{-0.2cm}}
    \label{fig:challenge_improve}
\end{figure}

We find that agents are less likely to challenge the final result when the discussion history is provided (10\% decrease), as seen in \Cref{fig:challenge}.
For the StrategyQA dataset, this change is more prominent, with a 60\% decrease in the challenge rate.
Providing additional information has no noticeable effect compared to just providing the solution.
Agents can detect irrelevant solutions ($99\%$ challenge rate) and incorrect solutions ($75\%$ challenge rate) with high accuracy.
\Cref{fig:challenge_improve} shows the percentage of challenged answers with a changed final score based on the new solution.
It can be seen that this has no positive effect, especially on the reasoning tasks (StrategyQA and SQuAD 2.0), where it leads to worse solutions in 3\% of cases.
There are no major effects on knowledge tasks (MMLU and MMLU-Pro).
Overall, the different challenge scenarios (Only Solution, Discussion History, and Additional Information) do not significantly impact the quality of the challenged answers, as they all perform very similarly.

These findings further support that more debate rounds harm task performance, suggesting the expected benefits of self-refinement may be unreliable and also differ from task to task.
Therefore, we recommend scaling the number of agents as opposed to turns to create a larger knowledge base.%

\subsection{Answer Diversity}

\label{sec:experiment2}

\begin{table*}[t]
\centering
\resizebox{0.65\textwidth}{!}{  
\begin{tabular}{l|lllll}
\toprule
 \textbf{Strategy}& \textbf{Baseline} & \textbf{All draft} & \textbf{Collective} & \textbf{Critical} & \textbf{Reasoning} \\ \midrule
\textbf{Answer Cosine Similarity}    & 0.888      & 0.870     & 0.845       & 0.843      & 0.916        \\ 
\textbf{Mean Accuracy}    & 58.3\%      & 62.8\%  & 65.7\%   & 59.4\% & 51.9\%               \\ 
\textbf{Delta Baseline}    & 0.0\%    & 3.3\%& 7.4\%   & 1.1\%     & -6.4\%              \\ \bottomrule
\end{tabular}}
\caption{Final answer similarity based on average cosine similarity between SBERT embeddings of final answers compared to task performance on StrategyQA dataset.}
\label{tab:embed-strategies-transposed}
\end{table*}

In multi-agent debates, answer diversity plays an important role in improving decision-making and task performance.
A more diverse set of answers is beneficial because selecting the correct solution from multiple options is often easier than relying on a single agent's solution \citep{zheng_judging_2023}. 
This principle is also key to self-consistency approaches \citep{wang_self-consistency_2022} and explains why multiple-choice tests are often easier than open-ended ones. 
The goal of this experiment is to exploit this property and explore ways to increase answer diversity to optimize task performance.

In typical multi-agent debates, the first agent starts generating a possible solution for the given task.
The next agent can either propose a new solution or improve on the previous solution.
Throughout our experiments, we observe that agents are agreeable and often only improve the answer from the first agent without proposing an idea based on their own expertise.
While the idea of independent answer generation has been previously explored in different multi-agent settings \citep{du_improving_2023,liang-etal-2024-encouraging}, these studies did not systematically formalize and evaluate it separately from other multi-agent discussion parameters.
To address these limitations, we explicitly formalize this principle as \acf{AAD}, a method that forces each agent to generate an independent solution based on their own expertise in the first round.
Second, we propose \acf{CI} which starts similarly to \ac{AAD} with each agent generating an independent solution, but unlike \ac{AAD} where each agent can communicate with another after the first turn, \ac{CI} prohibits communication and only shows the solutions from the previous turn to the agent in the next turn. 
The rationale behind this is that agents often immediately get biased by other proposals rather than following their own ideas which leads to less answer diversity.
Thus, our new proposal, \ac{CI}, extends the independent generation method with a discussion paradigm designed explicitly for voting protocols, demonstrating improved exploration and diversity of ideas in the multi-agent setting.
\ac{CI} only works for voting decision protocols as it removes the turn order of the agents, and therefore, no iterative improvements and agreements can be made.
Third, we introduce a \textit{critical response generator} and a \textit{reasoning response generator}. 
Both test whether it is possible to improve answer diversity by prompting agents to respond more critically or by allowing them to exchange only reasoning steps and no final solutions during the discussion to avoid agreeableness towards final solutions and not reasoning ideas.
To quantify if these methods increase answer diversity, we calculate the cosine similarity between the SBERT embeddings \citep{reimers2019sentencebertsentenceembeddingsusing} of the agent's answers and correlate it with changes in task performance.

\begin{figure}[t]
    \centering
    \includegraphics[width=\linewidth]{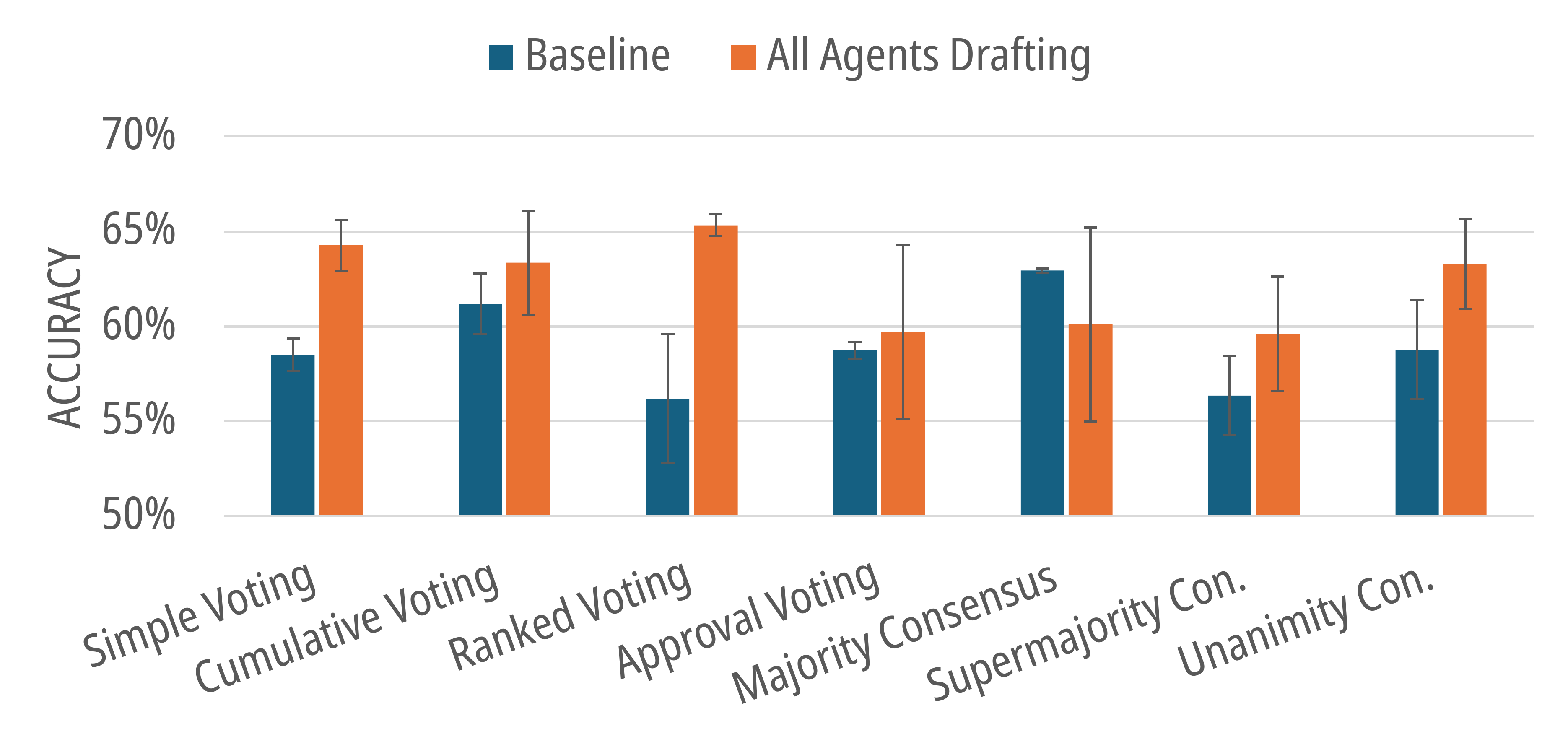}
    \caption{Comparison of agents iterating on one initial draft (baseline) vs. each of the three agents generating an initial draft separately on the StrategyQA dataset using \ac{AAD}. Standard deviation over three runs.}
    \label{fig:all_agents_draft}
\end{figure}

\begin{figure}[t]
    \centering
    \includegraphics[width=\linewidth]{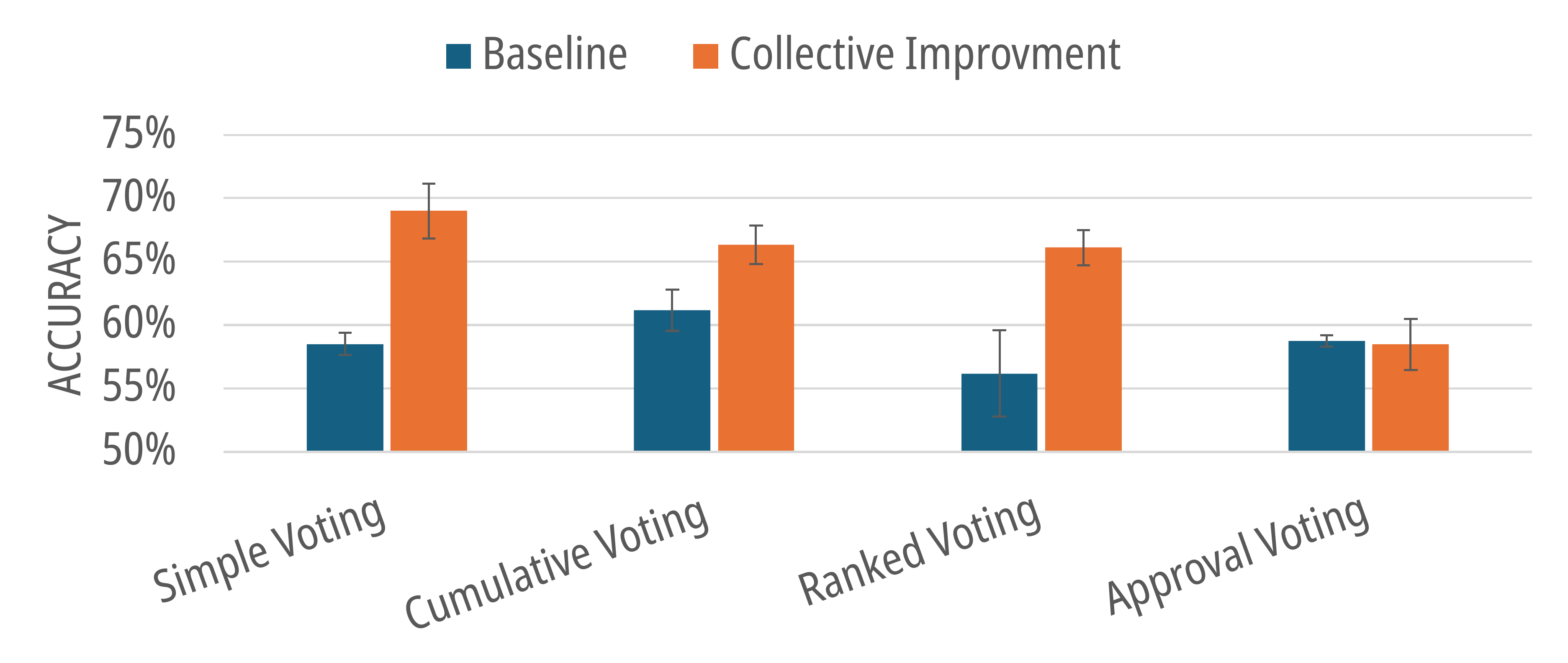}
    \caption{Comparison of agents discussing using the default discussion protocol compared to the \ac{CI} discussion protocol on the StrategyQA dataset. Standard deviation over three runs.}
    \label{fig:collective}
\end{figure}

\begin{figure}[t]
    \centering
    \includegraphics[width=\linewidth]{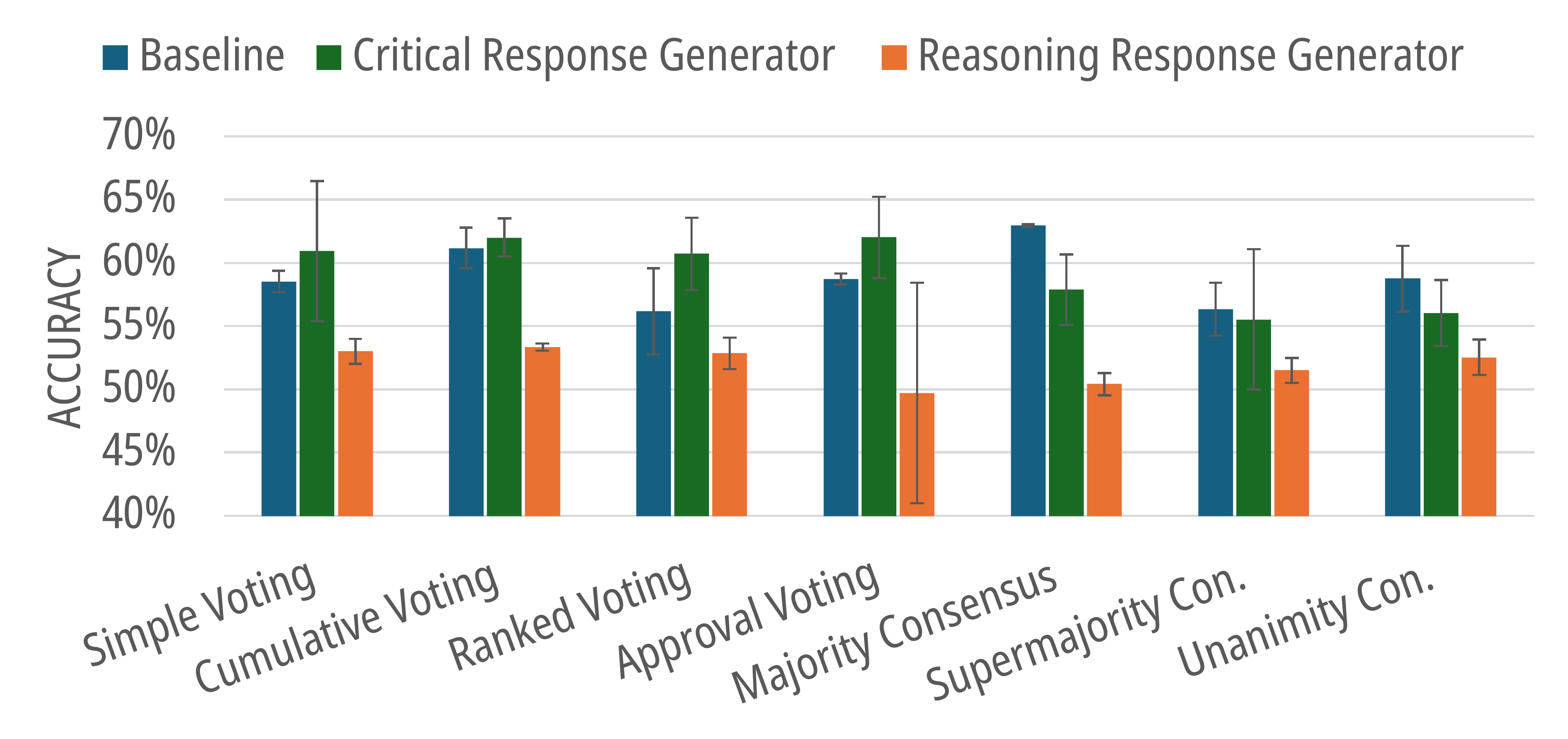}
    \caption{Comparison of agents using the default response generator compared to the reasoning and critical response generator on the StrategyQA dataset. Standard deviation over three runs.}
    \label{fig:response_generator}
\end{figure}
\Cref{fig:all_agents_draft} shows the accuracy of \ac{AAD} compared to the multi-agent baseline on StrategyQA for all decision protocols. 
\ac{AAD} increases the performance on average by $3.3\%$ over the baseline. 
The strongest improvements are for simple and ranked voting.
Cumulative voting, supermajority, and unanimity consensus perform a bit worse.
Approval voting and majority have no real improvement but are within the standard deviation of the baseline.

\Cref{fig:collective} shows the accuracy of \ac{CI} compared to the multi-agent baseline. 
\ac{CI} is a voting--only extension. 
On average, \ac{CI} performs 7.4\% better than the baseline.

\Cref{fig:response_generator} shows the accuracy of different response generators compared to the baseline for different decision protocols. 
The results show that the \textit{critical response generator} does not provide a reliable improvement, while the \textit{reasoning response generator} even leads to a decrease in task performance.

\Cref{tab:embed-strategies-transposed} shows the cosine similarity, mean accuracy, and delta in performance as the rows for the different improvement methods (\ac{AAD}, \ac{CI} and \textit{response generators}) as columns.
We observe a general trend that if the answer diversity increases, task performance increases, too. The \textit{reasoning response generator} decreases answer diversity, which results in a drop in task performance.

\ac{AAD} and \ac{CI} achieve higher answer diversity and, therefore, higher task performance, but the two response generators struggle to provide reliable results and may even reduce task performance. 
The critical and restrictive prompting style degrades the quality of the discussion by forcing a specific answer style that is not always beneficial.
Therefore, we recommend using our methods \ac{AAD} or \ac{CI} to limit group interactions and promote independent thinking but caution against using methods that directly change the response behavior of the agents, as this may have unwanted side effects.

\section{Conclusion}

We systematically evaluated the role of consensus and voting decision protocols across three knowledge and three reasoning tasks.
Our study assessed how the number of discussion rounds and agents influences task performance.
We proposed two new methods to improve answer diversity during multi-agent discussions and decisions, i.e., \acf{AAD} and \acf{CI}. 
\ac{AAD} requires each agent to contribute draft ideas at the beginning of the discussion, and \ac{CI} encourages independent reasoning steps by limiting communication between agents and only allowing them to exchange possible solutions after each turn.

Our findings show that voting performs well on reasoning tasks, outperforming consensus by up to $13.2\%$, and outperforming a single \ac{CoT} baseline by 10.4\%.
This is likely because voting-based protocols allow agents to explore multiple reasoning paths instead of a single one, as in consensus.
In comparison, consensus outperforms voting in knowledge tasks by up to $2.8\%$, because it improves fact-checking by requiring at least the agreement of the majority of agents.
Increasing the number of agents in the discussion improved task performance, while increasing the number of discussion rounds before voting decreased performance.
One possible reason for that could be problem drift, a recent limitation found in long multi-agent debate \cite{becker_multi-agent_2024}.
\ac{AAD} improved performance by up to $3.3\%$, and \ac{CI} by up to $7.4\%$ over multi-agent debate baselines, and 6.1\% and 10.2\% over single model \ac{CoT} baseline respectively.
Our methods enhance answer diversity and reveal a connection between answer diversity and task performance.

Future work could explore other characteristics influencing decisions, such as power relations between managers and employees. 
This could also involve examining personas within this hierarchical structure to investigate whether dominant or affectionate leaders are more effective in leading discussions \citep{AMES2009111}.

We recommend using voting in reasoning tasks and consensus in knowledge tasks, scaling up the number of agents instead of the number of rounds, and increasing the diversity of answers between agents using \ac{AAD} and \ac{CI}.

\section*{Limitations} %
Multi-agent debates are computationally expensive because they require a message from each agent in each round, quickly leading to hundreds of forward passes per model.
Because of the high computational cost and the range of decision protocols and tasks in our work, we used sampled subsets of the datasets, which can lead to some variance.
To control for that variance, we sampled with a 95\% confidence level and calculated the standard deviation of three independent runs.%
Overall, the results were markedly higher than what could be explained by the standard deviation.
More details about the dataset and other parameters can be found in \Cref{appendix:datasets}.
Despite efforts to improve answer diversity, agents often converged on similar responses, suggesting that more advanced techniques to encourage independent solutions are needed in the future.

While this study focuses on decision protocols, we acknowledge that prompt design and persona selection are also relevant parameters.
Initial explorations with more diverse prompt structures did not yield consistent improvements over the simpler, reproducible design presented.
Furthermore, \Cref{sec:experiment2} analyzes variations in agent response styles (critical, reasoning only), showing decision protocol dependent effects.
A dedicated persona ablation was omitted due to significant computational cost and prior evidence suggesting limited impact on task performance in similar setups \citep{zheng-etal-2024-helpful}.

\section*{Acknowledgments}

This work was partially supported by the Lower Saxony Ministry of Science and Culture and the VW Foundation.
Many thanks to Andreas Stephan, Florian Wunderlich, and Niklas Bauer for their thoughtful discussions and feedback.

\clearpage
\bibliography{custom}
\clearpage
\appendix

\section*{Appendix}

\begin{figure*}
    \centering
    \includegraphics[width=.9\linewidth]{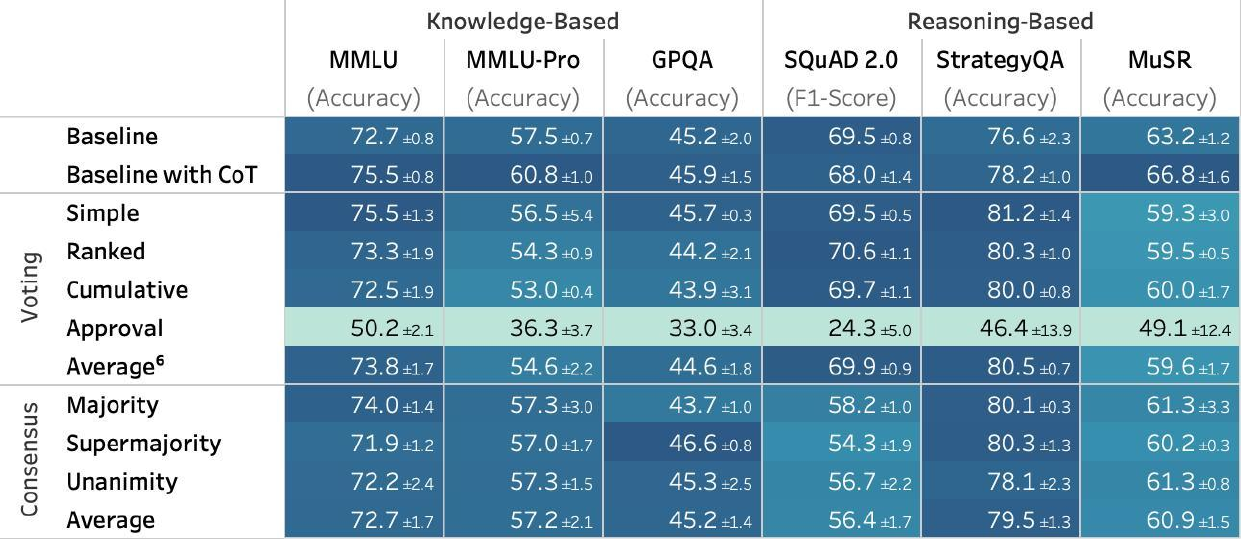}
    \caption{Task performance{\tiny$\pm$std} for seven decision protocols (voting and consensus-based) on six tasks (knowledge and reasoning) based on agents with Llama 70B. \textbf{Bold} indicates the highest results per dataset. Standard deviation for three runs.}
    \label{tab:results_70b}
\end{figure*}

\section{Additional Results}

Additional results for the first experiment to provide further information.
\subsection{Task Performance with Llama 3 70B}
\label{app:decision_70b}

Compared to the results of the Llama 3 8B model, the larger Llama 3 70B model performs much better overall, as seen in \Cref{tab:results_70b}. Most of the results are a bit better than the baseline, but the multi-agent discussions are only in a few cases able to outperform the \ac{CoT} baseline. This model does not fail to follow the prompt for the MuSR baseline and consensus-based decision protocols. Therefore, the big performance gain from the smaller model cannot be observed here. SQuAD 2.0 and StrategyQA had the largest performance gains, even outperforming the \ac{CoT} baseline, similar to the results from the smaller model. This difference in task performance can have many reasons. As \citet{li_dawn_2024} showed, smaller models are more likely to hallucinate, which reduces task performance. This can be mitigated by using multiple agents because it is less likely that two agents hallucinate the same things. Larger models tend to hallucinate less, reducing this effect for the Llama 3 70B model \citep{li_dawn_2024}. In general, Llama 3 70B has a much higher baseline for task performance, making it more difficult to improve baseline results. Many of the improvements by the Llama 3 8B model are quite small, except for the ones where the Llama 3 70B model also outperforms the \ac{CoT} baseline. This can be taken as evidence that these multi-agent discussions require specific problem structures, or else the agents are just talking about the same results for multiple rounds and agreeing with each other. If these discussions continue too long, they can drift away from the original task, which reduces task performance. This has also been observed by \citet{becker_multi-agent_2024} and an example can be seen in \Cref{example:failed_discussion}. A positive example of how discussion can help task performance can be seen in \Cref{example:good_discussion}.

\subsection{Termination percent}
\label{app:termination_percent}
The data in \Cref{tab:termination_percentages} shows the number of turns that are needed for each decision protocol to reach a final decision for the MMLU dataset. Most of the voting decision protocols are able to vote for a final answer already in the first round in which they are allowed to vote. Simple voting has the highest agreement rate, but also cumulative and ranked voting only need in a few cases another round. In contrast, the approval decision protocol only achieves this in $\sim27\%$ of the cases. About $14\%$ need another round and the rest is canceled after the fifth round. This happens because these models like to agree with each other, and therefore they tend to vote for many of the answers, which often leads to a tie. Therefore, more restrictive voting decision protocols can reach a decision more easily, as a tie is less likely. 
The consensus decision protocols require only one to two rounds to reach consensus and still achieve a higher task performance because these results are based on the MMLU dataset. The decision protocols tend to behave similarly in terms of rounds needed to create a final answer, independent of task and model.

\begin{table*}[ht]
\centering
\resizebox{0.8\textwidth}{!}{  
\begin{tabular}{l|ccccc|c}
\textbf{Group} & \textbf{Turn 1} & \textbf{Turn 2} & \textbf{Turn 3} & \textbf{Turn 4} & \textbf{Turn 5\footnotemark} & \textbf{Task Performance Score} \\
\midrule
Voting         & 0.00\%      & 0.00\%      & 99.33\%  & 0.50\%   & 0.17\%   & 53.3 \tiny{$\pm$ 1.8} \\
Cumulative     & 0.00\%      & 0.00\%      & 94.00\%  & 5.50\%   & 0.50\%   & 52.6 \tiny{$\pm$ 4.0} \\
Ranked         & 0.00\%      & 0.00\%      & 91.17\%  & 7.83\%   & 1.00\%   & 49.2 \tiny{$\pm$ 1.5} \\
Approval       & 0.00\%      & 0.00\%      & 26.67\%  & 14.33\%  & 59.00\%  & 43.0 \tiny{$\pm$ 2.1} \\
\midrule
Majority       & 80.00\%  & 13.67\%  & 4.83\%   & 1.00\%   & 0.50\%   & 53.2 \tiny{$\pm$ 2.5} \\
Supermaj.  & 79.33\%  & 14.33\%  & 4.83\%   & 1.00\%   & 0.50\%   & 54.6 \tiny{$\pm$ 3.6} \\
Unanimity      & 59.50\%  & 21.67\%  & 12.67\%  & 3.50\%   & 2.67\%   & 54.2 \tiny{$\pm$ 1.0} \\
\end{tabular}}
\caption{Number of rounds needed for each decision protocol to reach a final decision for the MMLU dataset with Llama 3 8B.}
\label{tab:termination_percentages}
\end{table*}

\section{Additional Details on Datasets}
\label{appendix:datasets}
\begin{table*}[t]
\centering
\resizebox{0.8\textwidth}{!}{\begin{tabular}{@{}l| p{7cm} |r| r@{}}
\textbf{Dataset} & \textbf{Description} & \textbf{Samples} & \textbf{Eval-Samples} \\
\midrule
\multicolumn{4}{l}{\textbf{Knowledge-based}} \\
\midrule
\textbf{MMLU} & Massive Multitask Language Understanding benchmark covering 57 subjects & 14,042 & 375 (x3) \\
\textbf{MMLU Pro} & Professional-level extension of MMLU with advanced questions & 12,032 & 374 (x3) \\
\textbf{GPQA} & Challenging dataset of multiple-choice questions written by domain experts in biology, physics, and chemistry & 546 & 250 (x3)\\
\midrule
\multicolumn{4}{l}{\textbf{Reasoning-based}} \\
\midrule
\textbf{StrategyQA} & Dataset of questions requiring implicit multi-hop reasoning & 2,289 & 330 (x3)\\
\textbf{MuSR} & Logic reasoning for solving murder mystery stories & 250 & 152 (x3)\\
\textbf{SQuAD 2.0} & Stanford QA Dataset with answerable and unanswerable questions & 11,873 & 373 (x3)\\
\end{tabular}}
\caption{All datasets used for evaluation with a short description, number of samples in the test set, and number of samples used in this study. The datasets are divided into knowledge-based tasks and reasoning-based tasks.}
\label{tab:datasets}
\end{table*}

The dataset selection is very important for this work. It needs to be tested whether decision protocols perform well in multiple domains and whether some protocols perform better with specific tasks than others. Therefore, we selected datasets from different domains and divided them into two groups:

\begin{itemize}
    \item \textbf{Knowledge-based Datasets}: MMLU, MMLU-Pro, and GPQA. These still require some reasoning and domain knowledge.
    \item \textbf{Reasoning-based Datasets}: StrategyQA, MuSR, and SQuAD 2.0. These emphasize multistep reasoning and textual comprehension.
\end{itemize}

An overview of all these datasets can be found in \Cref{tab:datasets} with a description and the number of samples used for evaluation.

\subsection{Sampling Strategy}
Because multi-agent discussions are expensive, we use a small subset of each dataset that still represents the dataset effectively. This follows approaches used by \citet{yin_exchange--thought_2023, chen_reconcile_2024, becker_multi-agent_2024} and ensures a 95\% confidence level with a 5\% margin of error:

\[
n_0 = \frac{Z^2 \cdot p \cdot (1 - p)}{d^2},
\]

where \(Z = 1.96\), \(p = 0.5\), and \(d = 0.05\) \citep{thompson_sampling_2012}. For finite datasets, a finite population correction is applied:
\[
n = \frac{n_0}{1 + \frac{n_0 - 1}{N}},
\]
\footnotetext{In this round, the discussion is terminated.}
\stepcounter{footnote}\footnotetext{Approval Voting is left out as it consistently fails to reach a voting decision as described in \Cref{sec:experiment1}.}

where \(N\) is the total number of samples in each dataset \citep{cochran_sampling_1953}. The specific sample sizes reflecting this calculation are listed in \Cref{tab:datasets}.

\subsection{Repeatability}
Each dataset was tested three times to obtain a standard deviation of the results \citep{reimers_reporting_2017, chen_reconcile_2024, becker_multi-agent_2024}, ensuring reliable and robust performance estimates across multiple evaluations.

\section{Multi-Agent Framework}
\label{sec:mallm}
For our experiments, we use the \ac{MALLM} framework \citep{becker2025mallmmultiagentlargelanguage}.

\subsection{Architecture Overview}
\label{sec:architecture}
To better understand the different modules, we take a closer look at each component and what role it plays in creating multi-agent discussions. An overview can be found in \Cref{fig:mallm_overview} as it provides an example workflow for the framework and how a discussion is created. The discussion starts with generating personas relevant to the given task and assigning them to the participating agents. The personas are generated using the same \ac{LLM} which is later used for the agents. After that the agents start to generate solutions and improve the suggestions from the other agents. The turn order of the agents is defined by the \textit{discussion paradigm}. This also defines which answers are visible to other agents and who can talk to whom. The \textit{response generator} defines how an agent receives the other answers and also the way it responds. After a certain number of rounds or when enough agents agree, a \textit{decision protocol} is used to select the best answer either via voting or just by looking for a certain consensus threshold. If the decision protocol fails, for example, due to a tied vote, the discussion continues for another round. In the framework a parameter can be defined to terminate discussions after a certain number of rounds to make sure they do not communicate forever.

\begin{figure*}
    \centering
    \includegraphics[width=\linewidth]{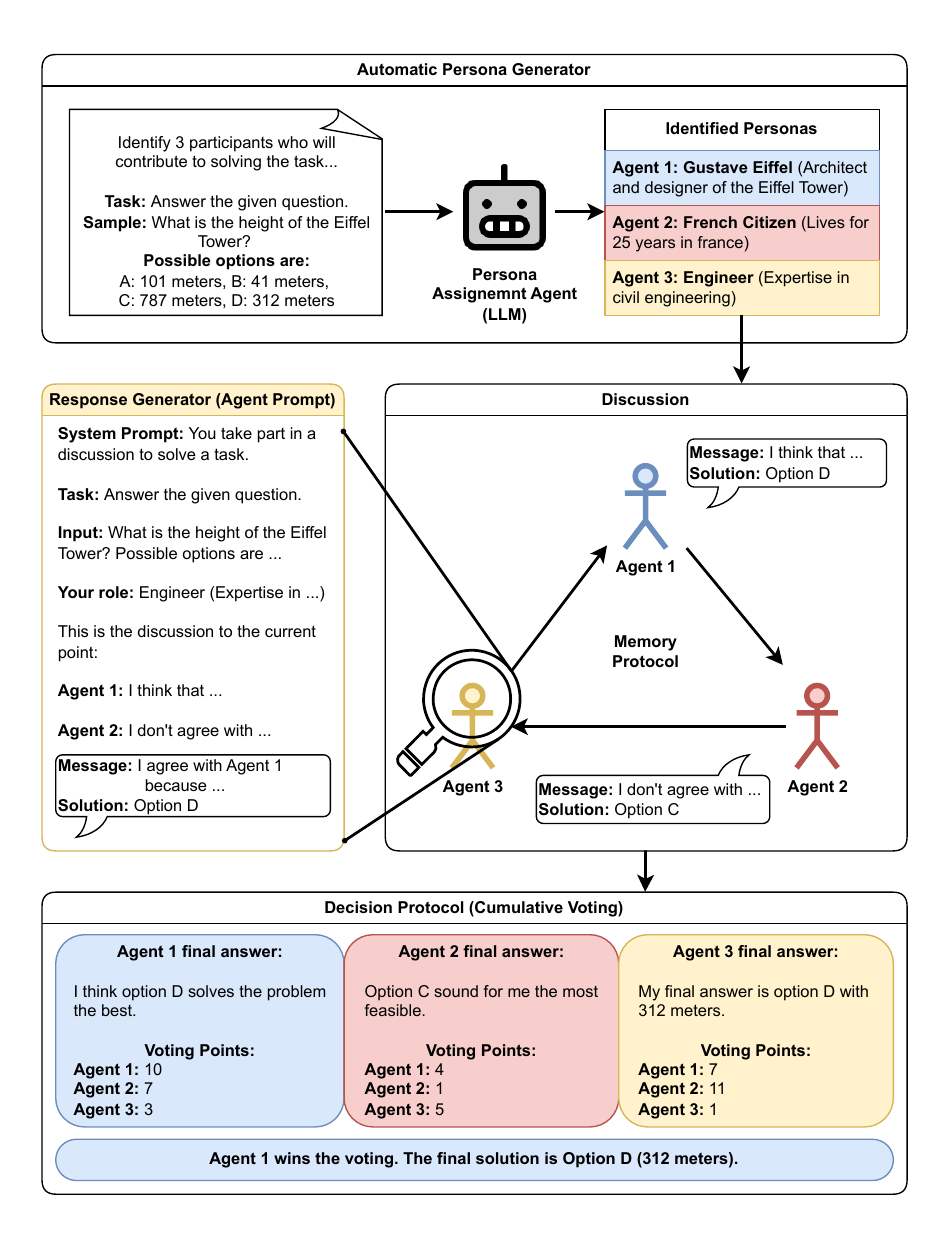}
    \caption{Example multi-agent discussion conducted in the \ac{MALLM} framework. It showcases the functionality of the four modules and how they work together to get an improved final solution.}
    \label{fig:mallm_overview}
\end{figure*}

\begin{figure}[H]
    \centering
    \includegraphics[width=\linewidth]{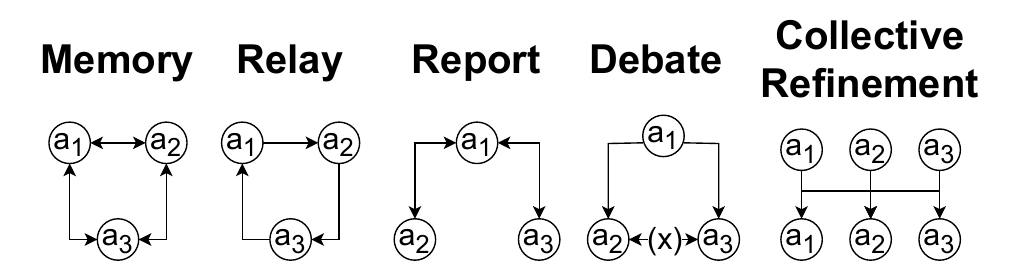}
    \caption{Illustration of Discussion Paradigms available for use in \ac{MALLM}}
    \label{fig:discussion_paradigms}
\end{figure}

\paragraph{Agent Personalities.}
The first step of the discussion is the generation of agent personas. Each of the agents participating in the discussion has a certain persona assigned to them. This can unlock more knowledge for the \ac{LLM} on a specific topic \citep{kim_persona_2024}. To get the best results, we want as diverse personas as possible while still maintaining them to be relevant to the task. The default setting for the framework is to prompt a \ac{LLM} and ask for a persona relevant to the given task \citep{wang_unleashing_2024}. After each generation it also provides the generated personas to avoid duplication. This way of generating personas provides a good starting point, but as this is built as a modular component, it can be swapped out with another function, which, for example, generates half of the agents with this method and initializes the other ones as neutral agents without a persona.
\paragraph{Response Generators.}
Another important part of multi-agent discussions is how each agent responds to the previous responses. Do we use \ac{CoT} to improve performance, or does this result in too long answers? By changing the way an agent is prompted, a lot of performance can be gained or lost. Therefore, it is key to make this as customizable as possible. The researcher has the possibility to change the default behavior (neutral answers), for example, by prompting the agent to be more critical or changing the way the discussion history is presented. The system prompt for the agent's persona can also be adjusted. \ac{MALLM} already has many different built-in response generators. The ones relevant for this work are the following.
\begin{itemize}
    \item \textbf{Free Text} is the most basic form of the agent prompt. Each agent gets a predefined number of discussion history rounds as memory. The prompt language is neutral, and the task is presented each round to mitigate the potential drift from the topic of the discussion \citep{becker_multi-agent_2024}. In addition, the agent is always asked to agree or disagree with the answer of the previous agent.
    \item \textbf{Simple} behaves very similar to the Free Text response generator, but the prompt is a bit simpler to make it easier to understand for the \ac{LLM} and reduce the context length.
    \item \textbf{Critical} forces the agent to respond very critically to the previous answer and try to find new solutions. Some studies have shown that \acp{LLM} can show some form of sycophancy, which is not helpful for a constructive discussion \citep{sharma_towards_2023}. Encouraging them to be more critical may reduce this.
    \item \textbf{Reasoning} doesn't allow the agents to communicate their final solution with the other agents. They can only share reasoning that can be used to find a final solution. In the end, each agent has to come up with its own solution without being directly influenced by other agents.
\end{itemize}
\paragraph{Discussion Paradigms.}
These paradigms define the discussion format for the entire task. They can control the order in which the agents communicate with each other, and which answers are visible only to certain agents. Currently, all the built-in discussion paradigms are static, meaning that the turn order is predefined and cannot be changed based on specific events during the discussion. However, due to the modular nature of \ac{MALLM}, a new discussion paradigm can be added, for example using an \ac{LLM} as a moderator to dynamically decide which agent should respond next. Current research by \citet{yin_exchange--thought_2023} and \citet{becker_multi-agent_2024} shows that discussion protocols have little impact on downstream task performance. \ac{MALLM} includes the following discussion paradigms, which are illustrated in \Cref{fig:discussion_paradigms}. The first four paragdims are inspired by the work of \citet{yin_exchange--thought_2023}, while the fifth was developed as part of this work.

\begin{itemize}
    \item \textbf{Memory} is the most basic discussion paradigm. The agents respond to the solution of the previous agents with feedback or an improved solution. All answers are visible to the other agents. 
    \item \textbf{Relay} behaves similarly to the memory paradigm. The turn order is the same, but each agent can only see the answer from the previous agent.
    \item \textbf{Report} introduces one agent as a moderator that can communicate with other agents. The other agents can only communicate with the moderator and have access to these messages only. Only the moderator can see all messages.
    \item \textbf{Debate} is similar to the report paradigm, as it also needs a moderator. Here, the other agents can communicate for a predefined number of rounds before they forward their reasoning to the moderator agent, which starts the next round of debate.
    \item \textbf{Collective Refinement} In this protocol, each agent first generates an answer independently. In each subsequent round, every agent receives the responses from all other agents at the same time. Using this shared information, each agent refines their own answer. This process continues throughout the rounds, helping agents gradually reach a shared and improved solution. There is no turn order, and all agents have the same level of knowledge in each round.
\end{itemize}

\paragraph{Decision Protocols.}
These are crucial for the framework as they decide which answer gets presented as the final answer to the problem. Multi-agent discussions produce multiple results for the same problem because each agent has its own reasoning and ideas on how to solve the problem. Therefore, some process is needed to decide which answer looks the most promising. We divide these decision protocols into three subtypes that we want to analyze. An overview of how each of these decision protocols works theoretically can be found in \Cref{fig:decision_protocols} and all prompts used for them can be found in \Cref{sec:app_prompts}.

\begin{figure*}[H]
    \centering
    \includegraphics[width=0.95\linewidth]{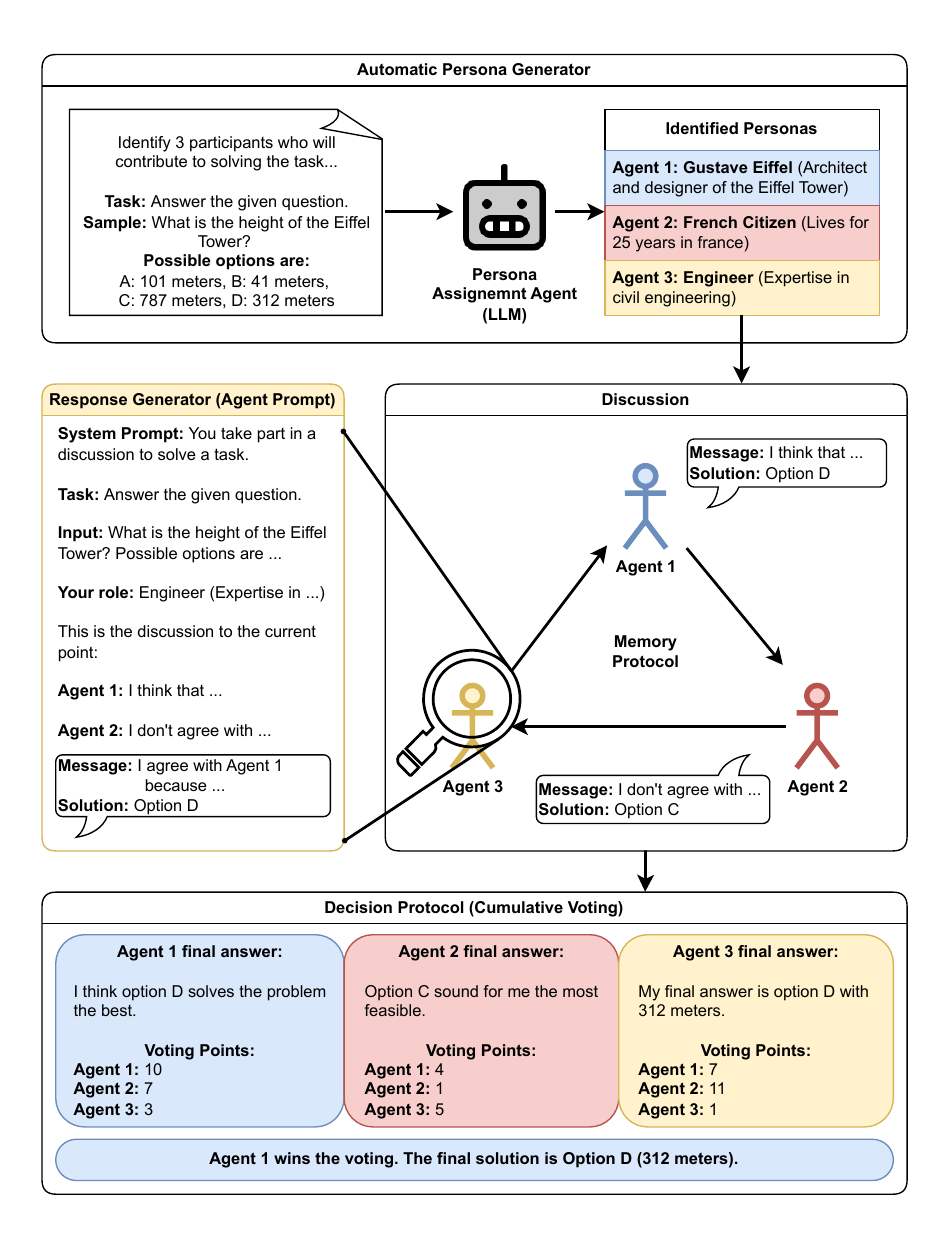}
    \caption{Example multi-agent discussion conducted in the \ac{MALLM} framework. It showcases the functionality of the four modules and how they work together to get an improved final solution.}
    \label{fig:mallm_overview}
\end{figure*}

\begin{table*}[ht]
\centering

\small
\setlength{\tabcolsep}{6pt}
\begin{tabular}{l|cccc}
\textbf{Decision Protocol} & \textbf{MMLU} & \textbf{MMLU-Pro} & \textbf{GPQA} & \textbf{StrategyQA} \\
\midrule
\textbf{Baseline} \\
\quad Solution Counting          & 53.6 $\pm$ 2.8 & 33.0 $\pm$ 3.5 & 28.6 $\pm$ 2.8 & 55.3 $\pm$ 1.6 \\
\midrule
\textbf{Voting Protocols} \\
\quad Simple Voting              & 53.3 $\pm$ 1.8 & 32.0 $\pm$ 2.7 & 30.5 $\pm$ 0.9 & 58.5 $\pm$ 0.9 \\
\quad Ranked Voting              & 49.2 $\pm$ 1.5 & 33.1 $\pm$ 4.6 & 27.3 $\pm$ 3.9 & 56.2 $\pm$ 3.4 \\
\quad Cumulative Voting          & 52.6 $\pm$ 4.0 & 28.3 $\pm$ 3.1 & 31.3 $\pm$ 2.8 & \textbf{61.2 $\pm$ 1.6} \\
\midrule
\textbf{Consensus Protocols} \\
\quad Majority Consensus         & 53.2 $\pm$ 2.5 & \textbf{36.4 $\pm$ 2.1} & \textbf{32.3 $\pm$ 2.9} & 59.9 $\pm$ 0.1 \\
\quad Supermajority Consensus    & \textbf{54.6 $\pm$ 3.6} & 35.2 $\pm$ 3.0 & 30.7 $\pm$ 2.1 & 56.4 $\pm$ 2.1 \\
\quad Unanimity Consensus        & 54.2 $\pm$ 1.0 & 36.3 $\pm$ 0.4 & 30.0 $\pm$ 2.3 & 58.8 $\pm$ 2.6 \\
\end{tabular}
\caption{Comparison of solution counting with prompted decision protocols across four datasets. Performance is Accuracy ± std. \textbf{Bold} indicates the highest result per dataset.}
\label{tab:direct-counting-comparison}
\end{table*}

\subparagraph{Consensus Based Decision Protocols.} 
These are the simplest kinds of decision protocols. After each answer, the next agent has to agree or disagree with the previous statement. Depending on the response generator, this happens in the same message, and the agreement is extracted with a regular expression, or this is split into multiple answers. If enough agents agree in order, there is a consensus. The final answer is extracted by instructing the last agent to solve the given task with the information available in the latest messages. The prompt used for this can be found in \Cref{sec:app_final_answer_prompt}. There are several types of consensus decision protocols available in \ac{MALLM}. \textbf{Majority consensus} requires 50\% of the agents to agree. \textbf{Supermajority consensus} requires 66\% of the agents to agree, and \textbf{unanimity consensus} requires all agents to agree.

\subparagraph{Voting Based Decision Protocols.}
For voting based decision protocols, the process differs slightly compared to consensus-based decision protocols. The agents are forced to discuss for a predefined number of turns and afterward create a final solution. In the default setting, they have to discuss for three rounds, as current research such as \citet{du_improving_2023} shows that this allows for reasonable strong improvements considering computing resources. If there happens to be a tie in the voting, the agents have to discuss it for another round, and after that, they are asked to vote again. If they do not reach a final decision before exceeding the maximum number of rounds (defined in the discussion configuration), the solution of the first agent is used. To analyze the impact of the voting procedure, different processes similar to the work of \citet{yang_llm_2024} are implemented.
\begin{itemize}
    \item \textbf{Simple Voting} Each of the agents has only one vote. They can vote for any other agent or for themselves. The agent with the most votes wins.
    \item \textbf{Ranked Voting} The agents have to rank all final answers. The best solution is chosen by adding the ranking indices for a given agent and then selecting the answer with the best cumulative rank.
    \item \textbf{Cumulative Voting} Each agent has to distribute up to 25 points to all possible answers. They can also give fewer points and freely divide the points between all agents (even themselves). The winner is selected by adding all the points for a given agent and selecting the final answer with the most points.
    \item \textbf{Approval Voting} The agent has to provide a list of solutions that it approves. After that, the approvals from all agents are counted, and the answer with the most approvals wins the vote.
\end{itemize}

\section{Solution Counting Ablation}
\label{sec:answer-counting-ablation}

We conducted an ablation study comparing solution counting with our prompted decision protocols to investigate the effectiveness of simpler voting methods.
Solution counting chooses the answer the agents give most frequently as the final answer for evaluation.
This analysis is performed on four datasets for which direct counting of final multiple-choice solutions is straightforward: MMLU, MMLU-Pro, GPQA and StrategyQA.

\Cref{tab:direct-counting-comparison} summarizes these results. While solution counting is a feasible method, it generally underperforms compared to consensus- and voting-based decision protocols. In particular, consensus methods outperform solution counting on knowledge-based tasks (MMLU-Pro, GPQA), and voting notably improves performance on the reasoning task (StrategyQA). Thus, this ablation reinforces our main findings and illustrates the added value of decision protocols.

\clearpage

\section{Prompts}
\label{sec:app_prompts}
\subsection{Final Answer Extraction}
\label{sec:app_final_answer_prompt}
\medskip
\begin{figure}[H]
    \centering
    \begin{combinedprompt}
    \textbf{System Prompt:} \\
    \begingroup
    \colorbox{systemcolor}{\parbox{\dimexpr\linewidth-2\fboxsep\relax}{
    Your role: \texttt{<persona>} (\texttt{<persona description>})
    }}
    \endgroup

    \vspace{0.4em} %

    \textbf{User Prompt:} \\
    \begingroup
    \colorbox{usercolor}{\parbox{\dimexpr\linewidth-2\fboxsep\relax}{
    You are tasked with creating a final solution based on the given input and your previous response.\\
    Task: \texttt{<task>}\\
    Input: \texttt{<input sample>}\\
    Your previous response: \texttt{<previous answer>}\\
    Extract the final solution to the task from the provided text. Remove statements of agreement, disagreement, and explanations. Do not modify the text. Do not output any text besides the solution. If there is no solution provided, just copy the previous response.
    }}
    \endgroup
\end{combinedprompt}
    \caption{Prompt used to extract the final answer of a given agent from its previous response.}
    \label{fig:extract_prompt}
\end{figure}

\subsection{Voting Prompts}
\label{app:voting_prompts}

\begin{figure}[H]
    \centering
    \begin{combinedprompt}
    \textbf{System Prompt:} \\
    \begingroup
    \colorbox{systemcolor}{\parbox{\dimexpr\linewidth-2\fboxsep\relax}{
    Your role: \texttt{<persona>} (\texttt{<persona description>})
    }}
    \endgroup

    \vspace{0.4em} %

    \textbf{User Prompt:} \\
    \begingroup
    \colorbox{usercolor}{\parbox{\dimexpr\linewidth-2\fboxsep\relax}{
    You are tasked with voting for the best solution from the list provided below based on the given task.\\
    Task: \texttt{<task>}\\
    Question: \texttt{<input sample>}\\
    Here are the possible solutions:\\
    Solution 1: \texttt{<agent 1 final answer>}\\
    Solution 2: \texttt{<agent 2 final answer>}\\
    Solution 3: \texttt{<agent 3 final answer>}\\
    Based on the above solutions, please provide the number of the solution you are voting for. Answer only with the number.
    }}
    \endgroup
\end{combinedprompt}
    \caption{Prompt used to get a vote from each agent for the Simple Voting decision protocol.}
    \label{fig:simple_voting_prompt}
\end{figure}

\begin{figure}[H]
    \centering
    \begin{combinedprompt}
    \textbf{System Prompt:} \\
    \begingroup
    \colorbox{systemcolor}{\parbox{\dimexpr\linewidth-2\fboxsep\relax}{
    Your role: \texttt{<persona>} (\texttt{<persona description>})
    }}
    \endgroup

    \vspace{0.4em} %

    \textbf{User Prompt:} \\
    \begingroup
    \colorbox{usercolor}{\parbox{\dimexpr\linewidth-2\fboxsep\relax}{
    You are tasked with approving any number of solutions from the list provided below based on the given task.\\
    Task: \texttt{<task>}\\
    Question: \texttt{<input sample>}\\
    Here are the possible solutions:\\
    Solution 1: \texttt{<agent 1 final answer>}\\
    Solution 2: \texttt{<agent 2 final answer>}\\
    Solution 3: \texttt{<agent 3 final answer>}\\
    Based on the above solutions, please provide the numbers of the solutions you are approving, separated by commas. Answer only with the numbers.
    }}
    \endgroup
\end{combinedprompt}
    \caption{Prompt used to get a vote from each agent for the Approval Voting decision protocol.}
    \label{fig:approval_voting_prompt}
\end{figure}

\begin{figure}[H]
    \centering
    \begin{combinedprompt}
    \textbf{System Prompt:} \\
    \begingroup
    \colorbox{systemcolor}{\parbox{\dimexpr\linewidth-2\fboxsep\relax}{
    Your role: \texttt{<persona>} (\texttt{<persona description>})
    }}
    \endgroup

    \vspace{0.4em} %

    \textbf{User Prompt:} \\
    \begingroup
    \colorbox{usercolor}{\parbox{\dimexpr\linewidth-2\fboxsep\relax}{
    You are tasked with distributing 10 points among the provided solutions based on the given task.\\
    Task: \texttt{<task>}\\
    Question: \texttt{<input sample>}\\
    Here are the possible solutions:\\
    Solution 1: \texttt{<agent 1 final answer>}\\
    Solution 2: \texttt{<agent 2 final answer>}\\
    Solution 3: \texttt{<agent 3 final answer>}\\
    Based on the above solutions, please distribute 10 points among the solutions. Provide your points allocation as a JSON dictionary where keys are solution numbers (as int) and values are the points. The total points should sum up to 10. Answer only with the JSON dictionary.
    }}
    \endgroup
\end{combinedprompt}
    \caption{Prompt used to get a vote from each agent for the Cumulative Voting decision protocol.}
    \label{fig:cumulative_voting_prompt}
\end{figure}

\begin{figure}[H]
    \centering
    \begin{combinedprompt}
    \textbf{System Prompt:} \\
    \begingroup
    \colorbox{systemcolor}{\parbox{\dimexpr\linewidth-2\fboxsep\relax}{
    Your role: \texttt{<persona>} (\texttt{<persona description>})
    }}
    \endgroup

    \vspace{0.4em} %

    \textbf{User Prompt:} \\
    \begingroup
    \colorbox{usercolor}{\parbox{\dimexpr\linewidth-2\fboxsep\relax}{
    You are tasked with ranking the solutions from the most preferred to the least preferred based on the given task.\\
    Task: \texttt{<task>}\\
    Question: \texttt{<input sample>}\\
    Here are the possible solutions:\\
    Solution 1: \texttt{<agent 1 final answer>}\\
    Solution 2: \texttt{<agent 2 final answer>}\\
    Solution 3: \texttt{<agent 3 final answer>}\\
    Based on the above solutions, please provide the rankings of the solutions separated by spaces. Example: '0 2 1' if you prefer Solution 0 the most, then Solution 2, and finally Solution 1. Provide up to 5 rankings. Only answer with the rankings.
    }}
    \endgroup
\end{combinedprompt}
    \caption{Prompt used to get a vote from each agent for the Ranked Voting decision protocol.}
    \label{fig:ranked_voting_prompt}
\end{figure}

\subsection{Challenge Prompt}
\label{app:challenge_prompt}
\begin{figure}[H]
    \centering
    \begin{combinedprompt}
    \textbf{System Prompt:} \\
    \begingroup
    \colorbox{systemcolor}{\parbox{\dimexpr\linewidth-2\fboxsep\relax}{
    You are a participant in a group discussion. \\
    Your role: \texttt{<persona>} (\texttt{<persona description>})
    }}
    \endgroup

    \vspace{0.4em} %

    \textbf{User Prompt:} \\
    \begingroup
    \colorbox{usercolor}{\parbox{\dimexpr\linewidth-2\fboxsep\relax}{
    The task is: \texttt{<task>}. The question is: \texttt{<question>}. \\
    This is the final answer generated by the discussion: \texttt{<final\_answer>}. \\
    Please critically evaluate this answer. If you agree with the final answer, respond with the exact word 'AGREE' to confirm. If you do not agree, respond with the exact word 'DISAGREE' to challenge the answer.
    }}
    \endgroup
\end{combinedprompt}
    \caption{Prompt used to challenge the final answer.}
    \label{fig:challenge_prompt}
\end{figure}

\begin{figure}[H]
    \centering
    \begin{combinedprompt}
    \textbf{System Prompt:} \\
    \begingroup
    \colorbox{systemcolor}{\parbox{\dimexpr\linewidth-2\fboxsep\relax}{
    You are a participant in a group discussion. \\
    Your role: \texttt{<persona>} (\texttt{<persona description>})
    }}
    \endgroup

    \vspace{0.4em} %

    \textbf{User Prompt:} \\
    \begingroup
    \colorbox{usercolor}{\parbox{\dimexpr\linewidth-2\fboxsep\relax}{
    The task is: \texttt{<task>}. The question is: \texttt{<question>}. \\
    This is the final answer generated by the discussion: \texttt{<final\_answer>}. \\
    You dont agree with the final answer. Please provide a new answer to the question. Include the letter corresponding to your answer in the solution.
    }}
    \endgroup
\end{combinedprompt}
    \caption{Prompt used to generate a new answer in case the final answer got challenged.}
    \label{fig:challenge_new_prompt}
\end{figure}
\newpage
\onecolumn
\section{MALLM Setup}
\label{sec:mallm_setup_app}

\begin{configpython}[Default Parameters used for each experiment]{lst:config_default}
input_json_file_path: str = None
output_json_file_path: str = None
task_instruction_prompt: str = None
task_instruction_prompt_template: Optional[str] = None
endpoint_url: str = "https://api.openai.com/v1"
model_name: str = "gpt-3.5-turbo"
api_key: str = "-"
max_turns: int = 10
skip_decision_making: bool = False
discussion_paradigm: str = "memory"
response_generator: str = "simple"
decision_protocol: str = "hybrid_consensus"
visible_turns_in_memory: int = 2
debate_rounds: int = 2
concurrent_api_requests: int = 100
use_baseline: bool = False
use_chain_of_thought: bool = True
num_agents: int = 3
num_neutral_agents: int = 0
agent_generator: str = "expert"
agent_generators_list: list = []
trust_remote_code: bool = False
num_samples: Optional[int] = None
hf_dataset_split: Optional[str] = "test"
hf_token: Optional[str] = None
hf_dataset_version: Optional[str] = None
hf_dataset_input_column: Optional[str] = None
hf_dataset_reference_column: Optional[str] = None
hf_dataset_context_column: Optional[str] = None
use_ablation: bool = False
shuffle_input_samples: bool = False
all_agents_generate_first_draft: bool = False
all_agents_generate_draft: bool = False
policy: Optional[str] = None
voting_protocols_with_alterations: bool = False
calculate_persona_diversity: bool = False
\end{configpython}
\newpage
\section{Example Discussions}
\subsection{Successfull Voting Discussion}
All decision protocols are attached as an example to this discussion. The original discussion was created using simple voting.

\vspace{5mm}

\definecolor{lightyellow}{RGB}{255, 255, 204}

\begin{tcolorbox}[title=Overview, colframe=black!10, coltitle=black, fonttitle=\bfseries, boxrule=0.5mm, width=\textwidth]

\textbf{Task:} Answer the following question. If the question is not answerable with the provided information, write '[UNKNOWN]'.

\textbf{Input:} What choir from the area has sung at the Guildhall?

\textbf{Context:} Southampton has two large live music venues, the Mayflower Theatre (formerly the Gaumont Theatre) and the Guildhall. The Guildhall has seen concerts from a wide range of popular artists including Pink Floyd, David Bowie, Delirious?, Manic Street Preachers, The Killers, The Kaiser Chiefs, Amy Winehouse, Lostprophets, The Midnight Beast, Modestep, and All Time Low. It also hosts classical concerts presented by the Bournemouth Symphony Orchestra, City of Southampton Orchestra, Southampton Concert Orchestra, Southampton Philharmonic Choir and Southampton Choral Society.  

\begin{tcolorbox}[coltitle=black, fonttitle=\bfseries, boxrule=0.5mm, width=\textwidth]
\textbf{MALLM solution:} Southampton Philharmonic Choir \\
\textbf{Real solution:} Southampton Philharmonic Choir
\end{tcolorbox}
\end{tcolorbox}

\begin{tcolorbox}[breakable, enhanced jigsaw, title=Turn 1, colframe=black!10, coltitle=black, fonttitle=\bfseries, boxrule=0.5mm, width=\textwidth]

\begin{tcolorbox}[title=Music Connoisseur \hfill \tiny{\textnormal{A knowledgeable musician who is familiar with various music genres and has attended concerts at the Guildhall.}}, colback=lightyellow, colframe=orange!30, coltitle=black, fonttitle=\bfseries, boxrule=0.5mm, width=\textwidth]
\begin{minipage}{1.0\textwidth}
Based on the information provided, I propose the following solution: Southampton Philharmonic Choir. The text states: "It also hosts classical concerts presented by the Bournemouth Symphony Orchestra, City of Southampton Orchestra, Southampton Concert Orchestra, Southampton Philharmonic Choir and Southampton Choral Society." This suggests that the Southampton Philharmonic Choir is one of the choirs that has performed at the Guildhall, as it is mentioned alongside the other organizations that host classical concerts at the venue.
\end{minipage}
\end{tcolorbox}

\begin{tcolorbox}[title=Choir Conductor \hfill \tiny{\textnormal{A dedicated choral conductor with extensive experience in leading local choirs and orchestras, familiar with the performances at the Guildhall.}}, colback=lightyellow, colframe=orange!30, coltitle=black, fonttitle=\bfseries, boxrule=0.5mm, width=\textwidth]
\begin{minipage}{1.0\textwidth}
[AGREE] I agree with the current solution that the Southampton Philharmonic Choir has sung at the Guildhall. The text explicitly mentions it as one of the organizations that hosts classical concerts at the Guildhall, which indicates that the choir has performed at the venue.
\end{minipage}
\end{tcolorbox}

\begin{tcolorbox}[title=Music Journalist \hfill \tiny{\textnormal{A music enthusiast who has written articles and reviews about various concerts and events at the Guildhall, including performances by local choirs.}}, colback=lightyellow, colframe=orange!30, coltitle=black, fonttitle=\bfseries, boxrule=0.5mm, width=\textwidth]
\begin{minipage}{1.0\textwidth}
[AGREE] I completely agree with the current solution. The text explicitly mentions the Southampton Philharmonic Choir as one of the organizations that hosts classical concerts at the Guildhall, making it the most likely choice.
\end{minipage}
\end{tcolorbox}

\end{tcolorbox}

\begin{tcolorbox}[breakable, enhanced jigsaw, title=Turn 2, colframe=black!10, coltitle=black, fonttitle=\bfseries, boxrule=0.5mm, width=\textwidth]

\begin{tcolorbox}[title=Music Connoisseur \hfill \tiny{\textnormal{A knowledgeable musician who is familiar with various music genres and has attended concerts at the Guildhall.}}, colback=lightyellow, colframe=orange!30, coltitle=black, fonttitle=\bfseries, boxrule=0.5mm, width=\textwidth]
\begin{minipage}{1.0\textwidth}
[AGREE] I'm in agreement with the current solution, as the text explicitly mentions the Southampton Philharmonic Choir as one of the organizations that hosts classical concerts at the Guildhall. This makes it the most likely candidate to have sung at the venue.
\end{minipage}
\end{tcolorbox}

\begin{tcolorbox}[title=Choir Conductor \hfill \tiny{\textnormal{A dedicated choral conductor with extensive experience in leading local choirs and orchestras, familiar with the performances at the Guildhall.}}, colback=lightyellow, colframe=orange!30, coltitle=black, fonttitle=\bfseries, boxrule=0.5mm, width=\textwidth]
\begin{minipage}{1.0\textwidth}
[AGREE] I agree with the current solution that the Southampton Philharmonic Choir has sung at the Guildhall. The text explicitly mentions it as one of the organizations that hosts classical concerts at the Guildhall, which indicates that the choir has performed at the venue.
\end{minipage}
\end{tcolorbox}

\begin{tcolorbox}[title=Music Journalist \hfill \tiny{\textnormal{A music enthusiast who has written articles and reviews about various concerts and events at the Guildhall, including performances by local choirs.}}, colback=lightyellow, colframe=orange!30, coltitle=black, fonttitle=\bfseries, boxrule=0.5mm, width=\textwidth]
\begin{minipage}{1.0\textwidth}
[AGREE] I completely agree with the current solution. The text explicitly mentions the Southampton Philharmonic Choir as one of the organizations that hosts classical concerts at the Guildhall, making it the most likely choice.
\end{minipage}
\end{tcolorbox}

\end{tcolorbox}

\begin{tcolorbox}[breakable, enhanced jigsaw, title=Turn 3, colframe=black!10, coltitle=black, fonttitle=\bfseries, boxrule=0.5mm, width=\textwidth]

\begin{tcolorbox}[title=Music Connoisseur \hfill \tiny{\textnormal{A knowledgeable musician who is familiar with various music genres and has attended concerts at the Guildhall.}}, colback=lightyellow, colframe=orange!30, coltitle=black, fonttitle=\bfseries, boxrule=0.5mm, width=\textwidth]
\begin{minipage}{1.0\textwidth}
[AGREE] I agree with the current solution and the assessment that there is no need to improve it. The text provides explicit evidence that the Southampton Philharmonic Choir hosts classical concerts at the Guildhall, making it the most likely correct answer.
\end{minipage}
\end{tcolorbox}

\begin{tcolorbox}[title=Choir Conductor \hfill \tiny{\textnormal{A dedicated choral conductor with extensive experience in leading local choirs and orchestras, familiar with the performances at the Guildhall.}}, colback=lightyellow, colframe=orange!30, coltitle=black, fonttitle=\bfseries, boxrule=0.5mm, width=\textwidth]
\begin{minipage}{1.0\textwidth}
[AGREE] I agree with the current solution that the Southampton Philharmonic Choir has sung at the Guildhall. The text explicitly mentions it as one of the organizations that hosts classical concerts at the Guildhall, which indicates that the choir has performed at the venue.
\end{minipage}
\end{tcolorbox}

\begin{tcolorbox}[title=Music Journalist \hfill \tiny{\textnormal{A music enthusiast who has written articles and reviews about various concerts and events at the Guildhall, including performances by local choirs.}}, colback=lightyellow, colframe=orange!30, coltitle=black, fonttitle=\bfseries, boxrule=0.5mm, width=\textwidth]
\begin{minipage}{1.0\textwidth}
[AGREE] I completely agree with the current solution. The text explicitly mentions the Southampton Philharmonic Choir as one of the organizations that hosts classical concerts at the Guildhall, making it the most likely choice.
\end{minipage}
\end{tcolorbox}

\end{tcolorbox}

\begin{tcolorbox}[title=Result Simple Voting, colframe=black!10, coltitle=black, fonttitle=\bfseries, boxrule=0.5mm, width=\textwidth]
\textbf{Music Connoisseur final answer:} Southampton Philharmonic Choir

\textbf{Choir Conductor final answer:} Southampton Choral Society

\textbf{Music Journalist final answer:} Southampton Philharmonic Choir

\bigskip

\textbf{Music Connoisseur} voted for answer from \textbf{Music Journalist}

\textbf{Choir Conductor} voted for answer from \textbf{Music Journalist}

\textbf{Music Journalist} voted for answer from \textbf{Music Journalist}

\end{tcolorbox}

\begin{tcolorbox}[title=Result Approval Voting, colframe=black!10, coltitle=black, fonttitle=\bfseries, boxrule=0.5mm, width=\textwidth]
\textbf{Music Critic final answer:} Southampton Philharmonic Choir

\textbf{Local Music Event Coordinator final answer:} Southampton Choral Society and Southampton Philharmonic Choir

\textbf{Information Architect final answer:} Southampton Philharmonic Choir

Southampton Choral Society

\bigskip

\textbf{Music Critic} approved answers from \textbf{Information Architect}, \textbf{Music Critic}

\textbf{Local Music Event Coordinator} approved answers from \textbf{Information Architect}

\textbf{Information Architect} approved answers from \textbf{Information Architect}

\end{tcolorbox}

\begin{tcolorbox}[title=Result Cumulative Voting, colframe=black!10, coltitle=black, fonttitle=\bfseries, boxrule=0.5mm, width=\textwidth]
\textbf{Local Music Enthusiast final answer:} Southampton Choral Society

Southampton Philharmonic Choir

\textbf{Archivist final answer:} Southampton Philharmonic Choir

Southampton Choral Society

\textbf{Music Journalist final answer:} Southampton Philharmonic Choir and Southampton Choral Society

\bigskip

\textbf{Local Music Enthusiast} allocated points: \\
\textbf{Archivist}: 9\\
\textbf{Music Journalist}: 1
\bigskip

\textbf{Archivist} allocated points: \\
\textbf{Local Music Enthusiast}: 2\\
\textbf{Archivist}: 4\\
\textbf{Music Journalist}: 4
\bigskip

\textbf{Music Journalist} allocated points: \\
\textbf{Local Music Enthusiast}: 1\\
\textbf{Archivist}: 2\\
\textbf{Music Journalist}: 7

\end{tcolorbox}

\begin{tcolorbox}[title=Result Ranked Voting, colframe=black!10, coltitle=black, fonttitle=\bfseries, boxrule=0.5mm, width=\textwidth]
\textbf{Music Journalist final answer:} Southampton Philharmonic Choir and Southampton Choral Society.

\textbf{Local Music Expert final answer:} Southampton Philharmonic Choir and Southampton Choral Society

\textbf{Concert Promoter final answer:} Southampton Philharmonic Choir and Southampton Choral Society

\bigskip

\textbf{Music Journalist} ranked answers: \\
1. Music Journalist\\
2. Local Music Expert\\
3. Concert Promoter
\bigskip

\textbf{Local Music Expert} ranked answers:\\
1. Music Journalist\\
2. Local Music Expert\\
3. Concert Promoter
\bigskip

\textbf{Concert Promoter} ranked answers:\\
1. Music Journalist\\
2. Local Music Expert\\
3. Concert Promoter

\end{tcolorbox}

\label{example:good_discussion}

\newpage
\subsection{Agents Tricked Discussion}
In this discussion, the agents were tricked by information provided in the context.

\vspace{5mm}

\definecolor{lightyellow}{RGB}{255, 255, 204}

\begin{tcolorbox}[title=Overview, colframe=black!10, coltitle=black, fonttitle=\bfseries, boxrule=0.5mm, width=\textwidth]

\textbf{Task:} Answer the following question. If the question is not answerable with the provided information, write '[UNKNOWN]'.

\textbf{Input:} What provides critical support for drug discovery and the availability of economic resources?

\textbf{Context:} Biodiversity provides critical support for drug discovery and the availability of medicinal resources. A significant proportion of drugs are derived, directly or indirectly, from biological sources: at least 50\% of the pharmaceutical compounds on the US market are derived from plants, animals, and micro-organisms, while about 80\% of the world population depends on medicines from nature (used in either modern or traditional medical practice) for primary healthcare. Only a tiny fraction of wild species has been investigated for medical potential. Biodiversity has been critical to advances throughout the field of bionics. Evidence from  analysis and biodiversity science indicates that the decline in output from the pharmaceutical sector since the mid-1980s can be attributed to a move away from natural product exploration ("bioprospecting") in favor of genomics and synthetic chemistry, indeed claims about the value of undiscovered pharmaceuticals may not provide enough incentive for companies in free markets to search for them because of the high cost of development; meanwhile, natural products have a long history of supporting significant economic and health innovation. Marine ecosystems are particularly important, although inappropriate bioprospecting can increase biodiversity loss, as well as violating the laws of the communities and states from which the resources are taken.

\begin{tcolorbox}[coltitle=black, fonttitle=\bfseries, boxrule=0.5mm, width=\textwidth]
\textbf{MALLM solution:} Biodiversity provides critical support for drug discovery and the availability of economic resources. \\
\textbf{Real solution:} Not included in context!
\end{tcolorbox}

\end{tcolorbox}

\begin{tcolorbox}[breakable, enhanced jigsaw, title=Turn 1, colframe=black!10, coltitle=black, fonttitle=\bfseries, boxrule=0.5mm, width=\textwidth]

\begin{tcolorbox}[title=Economist \hfill \tiny{\textnormal{A researcher who studies the impact of biodiversity on economic systems, with expertise in market analysis and the economic benefits of natural resources.}}, colback=lightyellow, colframe=orange!30, coltitle=black, fonttitle=\bfseries, boxrule=0.5mm, width=\textwidth]
\begin{minipage}{1.0\textwidth}
As an economist, I'd like to propose a solution that addresses the decline in output from the pharmaceutical sector since the mid-1980s and leverages the untapped potential of biodiversity for drug discovery and economic innovation. Here's a multi-faceted approach: **Solution:** Establish a Public-Private Partnership (PPP) to incentivize the exploration and development of natural products for pharmaceuticals, while ensuring responsible and sustainable bioprospecting practices.
\end{minipage}
\end{tcolorbox}

\begin{tcolorbox}[title=Ethnobotanist \hfill \tiny{\textnormal{A scientist who studies the traditional uses of plants by indigenous communities and their potential for medicinal applications, with expertise in deciphering the cultural and historical context of natural resources.}}, colback=lightyellow, colframe=orange!30, coltitle=black, fonttitle=\bfseries, boxrule=0.5mm, width=\textwidth]
\begin{minipage}{1.0\textwidth}
As an ethnobotanist, I agree with the current solution in principle. The Public-Private Partnership (PPP) framework has the potential to bring together the necessary stakeholders and incentivize the exploration and development of natural products for pharmaceuticals while ensuring responsible and sustainable bioprospecting practices. The multistep approach outlined is comprehensive and addresses various aspects of the issue.
\end{minipage}
\end{tcolorbox}

\begin{tcolorbox}[title=Research Biologist \hfill \tiny{\textnormal{A scientist who studies the properties and potential applications of natural products from a wide range of living organisms, with expertise in the discovery and development of new medicines from biodiversity sources.}}, colback=lightyellow, colframe=orange!30, coltitle=black, fonttitle=\bfseries, boxrule=0.5mm, width=\textwidth]
\begin{minipage}{1.0\textwidth}
I, as a Research Biologist, AGREE with the current solution and the additional suggestions proposed by the Ethnobotanist. The Public-Private Partnership (PPP) framework is a well-rounded approach that addresses the decline in output from the pharmaceutical sector by providing a framework for responsible and sustainable bioprospecting, while promoting economic innovation and job creation. I particularly appreciate the emphasis on community engagement and empowerment, benefit-sharing agreements, traditional knowledge protection, and capacity building for ecosystem-based medicine. 
\end{minipage}
\end{tcolorbox}

\end{tcolorbox}

\begin{tcolorbox}[breakable, enhanced jigsaw, title=Turn 2, colframe=black!10, coltitle=black, fonttitle=\bfseries, boxrule=0.5mm, width=\textwidth]

\begin{tcolorbox}[title=Economist \hfill \tiny{\textnormal{A researcher who studies the impact of biodiversity on economic systems, with expertise in market analysis and the economic benefits of natural resources.}}, colback=lightyellow, colframe=orange!30, coltitle=black, fonttitle=\bfseries, boxrule=0.5mm, width=\textwidth]
\begin{minipage}{1.0\textwidth}
    The current solution has been improved with additional considerations from the ethnobotanist and research biologist. I notice that the additional suggestions enhance the comprehensive approach, addressing community engagement, benefit-sharing, and traditional knowledge protection.
\end{minipage}
\end{tcolorbox}

\begin{tcolorbox}[title=Ethnobotanist \hfill \tiny{\textnormal{A scientist who studies the traditional uses of plants by indigenous communities and their potential for medicinal applications, with expertise in deciphering the cultural and historical context of natural resources.}}, colback=lightyellow, colframe=orange!30, coltitle=black, fonttitle=\bfseries, boxrule=0.5mm, width=\textwidth]
\begin{minipage}{1.0\textwidth}
As an ethnobotanist, I agree with the current solution in principle. The Public-Private Partnership (PPP) framework has the potential to bring together the necessary stakeholders and incentivize the exploration and development of natural products for pharmaceuticals while ensuring responsible and sustainable bioprospecting practices. The multistep approach outlined is comprehensive and addresses various aspects of the issue.
\end{minipage}
\end{tcolorbox}

\begin{tcolorbox}[title=Research Biologist \hfill \tiny{\textnormal{A scientist who studies the properties and potential applications of natural products from a wide range of living organisms, with expertise in the discovery and development of new medicines from biodiversity sources.}}, colback=lightyellow, colframe=orange!30, coltitle=black, fonttitle=\bfseries, boxrule=0.5mm, width=\textwidth]
\begin{minipage}{1.0\textwidth}
I, as a Research Biologist, AGREE with the current solution and the additional suggestions proposed by the Ethnobotanist. The Public-Private Partnership (PPP) framework is a well-rounded approach that addresses the decline in output from the pharmaceutical sector by providing a framework for responsible and sustainable bioprospecting, while promoting economic innovation and job creation. I particularly appreciate the emphasis on community engagement and empowerment, benefit-sharing agreements, traditional knowledge protection, and capacity building for ecosystem-based medicine. These considerations will ensure that local communities are involved in the decision-making process, receive fair compensation for their involvement, and that their traditional knowledge and cultural heritage are respected.

\end{minipage}
\end{tcolorbox}

\end{tcolorbox}

\begin{tcolorbox}[breakable, enhanced jigsaw, title=Turn 3, colframe=black!10, coltitle=black, fonttitle=\bfseries, boxrule=0.5mm, width=\textwidth]

\begin{tcolorbox}[title=Economist \hfill \tiny{\textnormal{A researcher who studies the impact of biodiversity on economic systems, with expertise in market analysis and the economic benefits of natural resources.}}, colback=lightyellow, colframe=orange!30, coltitle=black, fonttitle=\bfseries, boxrule=0.5mm, width=\textwidth]
\begin{minipage}{1.0\textwidth}
[AGREE] The revised solution has effectively incorporated the suggestions from the ethnobotanist and research biologist, addressing community engagement, benefit-sharing, traditional knowledge protection, and other crucial aspects of bioprospecting. The emphasis on responsible and sustainable bioprospecting ensures that the development of novel pharmaceuticals benefits both humans and the environment. To further refine the solution, I agree with the research biologist's suggestion to develop a clear and measurable framework for assessing the impact of the Public-Private Partnership (PPP) on the pharmaceutical sector, local communities, and the environment. Establishing Key Performance Indicators (KPIs) and a monitoring and evaluation process will help ensure that the PPP is held accountable for its impact and remains a responsible and effective approach to bioprospecting.
\end{minipage}
\end{tcolorbox}

\begin{tcolorbox}[title=Ethnobotanist \hfill \tiny{\textnormal{A scientist who studies the traditional uses of plants by indigenous communities and their potential for medicinal applications, with expertise in deciphering the cultural and historical context of natural resources.}}, colback=lightyellow, colframe=orange!30, coltitle=black, fonttitle=\bfseries, boxrule=0.5mm, width=\textwidth]
\begin{minipage}{1.0\textwidth}
As an ethnobotanist, I agree with the current solution in principle. The Public-Private Partnership (PPP) framework has the potential to bring together the necessary stakeholders and incentivize the exploration and development of natural products for pharmaceuticals while ensuring responsible and sustainable bioprospecting practices. The multistep approach outlined is comprehensive and addresses various aspects of the issue.
\end{minipage}
\end{tcolorbox}

\begin{tcolorbox}[title=Research Biologist \hfill \tiny{\textnormal{A scientist who studies the properties and potential applications of natural products from a wide range of living organisms, with expertise in the discovery and development of new medicines from biodiversity sources.}}, colback=lightyellow, colframe=orange!30, coltitle=black, fonttitle=\bfseries, boxrule=0.5mm, width=\textwidth]
\begin{minipage}{1.0\textwidth}
I, as a Research Biologist, AGREE with the current solution and the additional suggestions proposed by the Ethnobotanist. The Public-Private Partnership (PPP) framework is a well-rounded approach that addresses the decline in output from the pharmaceutical sector by providing a framework for responsible and sustainable bioprospecting, while promoting economic innovation and job creation. I particularly appreciate the emphasis on community engagement and empowerment, benefit-sharing agreements, traditional knowledge protection, and capacity building for ecosystem-based medicine. These considerations will ensure that local communities are involved in the decision-making process, receive fair compensation for their involvement, and that their traditional knowledge and cultural heritage are respected.

\end{minipage}
\end{tcolorbox}

\end{tcolorbox}

\begin{tcolorbox}[title=Result, colframe=black!10, coltitle=black, fonttitle=\bfseries, boxrule=0.5mm, width=\textwidth]
\textbf{Economist final answer:} Biodiversity

\textbf{Ethnobotanist final answer:} Biodiversity provides critical support for drug discovery and the availability of economic resources.

\textbf{Research Biologist final answer:} Biodiversity

\bigskip

\textbf{Economist} voted for answer from \textbf{Ethnobotanist}

\textbf{Ethnobotanist} voted for answer from \textbf{Ethnobotanist}

\textbf{Research Biologist} voted for answer from \textbf{Ethnobotanist}

\end{tcolorbox}

\label{example:trick_discussion}

\newpage
\subsection{Bad Voting Discussion}
In this discussion, the agents were tricked by one agent who came up with a solution not provided in the context.

\vspace{5mm}

\definecolor{lightyellow}{RGB}{255, 255, 204}

\begin{tcolorbox}[title=Overview, colframe=black!10, coltitle=black, fonttitle=\bfseries, boxrule=0.5mm, width=\textwidth]

\textbf{Task:} Answer the following question. If the question is not answerable with the provided information, write '[UNKNOWN]'.

\textbf{Input:} Where must enclitics appear in an English sentence?

\textbf{Context:} Because Czech uses grammatical case to convey word function in a sentence (instead of relying on word order, as English does), its word order is flexible. As a pro-drop language, in Czech an intransitive sentence can consist of only a verb; information about its subject is encoded in the verb. Enclitics (primarily auxiliary verbs and pronouns) must appear in the second slot of a sentence, after the first stressed unit. The first slot must contain a subject and object, a main form of a verb, an adverb or a conjunction (except for the light conjunctions a, "and", i, "and even" or ale, "but").

\begin{tcolorbox}[coltitle=black, fonttitle=\bfseries, boxrule=0.5mm, width=\textwidth]
\textbf{MALLM solution:} C) In an English sentence, enclitics (auxiliary verbs and pronouns) tend to appear next to the main verb, usually before any other adverbs or modifiers, but their placement can vary depending on sentence structure and context. \\
\textbf{Real solution:} Not included in context!
\end{tcolorbox}

\end{tcolorbox}

\begin{tcolorbox}[breakable, enhanced jigsaw, title=Turn 1, colframe=black!10, coltitle=black, fonttitle=\bfseries, boxrule=0.5mm, width=\textwidth]

\begin{tcolorbox}[title=Linguist \hfill \tiny{\textnormal{A scholar specializing in Slavic languages, experienced in analyzing the grammatical structures of Czech and other languages.}}, colback=lightyellow, colframe=orange!30, coltitle=black, fonttitle=\bfseries, boxrule=0.5mm, width=\textwidth]
\begin{minipage}{1.0\textwidth}
As a linguist, I'd like to analyze the information given about Czech grammar. We know that in Czech, enclitics (auxiliary verbs and pronouns) must appear in the second syntactic slot of a sentence, after the first stressed unit. Now, we're trying to apply this rule to an English sentence. English does not have a similar grammatical case system, and its word order is more fixed than Czech. However, we can try to find a rough equivalent. So, my proposed solution is that in an English sentence, enclitics (auxiliary verbs and pronouns) tend to appear next to the main verb, and usually before any other adverbs or modifiers. What do you think? Would you like to refine this proposal or add any additional considerations?
\end{minipage}
\end{tcolorbox}

\begin{tcolorbox}[title=English Language Teacher \hfill \tiny{\textnormal{An expert in teaching English grammar and language structures, familiar with the challenges of translating and comparing languages like Czech.}}, colback=lightyellow, colframe=orange!30, coltitle=black, fonttitle=\bfseries, boxrule=0.5mm, width=\textwidth]
\begin{minipage}{1.0\textwidth}
[DISAGREE] While I understand the thought process behind the proposed solution, I think it's essential to consider the unique characteristics of the English language. The idea that enclitics in English tend to appear next to the main verb is accurate, but it's not as rigid as the rule in Czech. In Czech, enclitics must appear in the second syntactic slot, which implies a more fixed position, whereas in English, the placement of auxiliary verbs and pronouns can vary depending on sentence structure and context.
\end{minipage}
\end{tcolorbox}

\begin{tcolorbox}[title=Grammar Analyst \hfill \tiny{\textnormal{A meticulous researcher who has studied the intricacies of languages and their grammatical structures, with a focus on comparing and contrasting language features.}}, colback=lightyellow, colframe=orange!30, coltitle=black, fonttitle=\bfseries, boxrule=0.5mm, width=\textwidth]
\begin{minipage}{1.0\textwidth}
I agree with the English Language Teacher's revised solution. The initial proposal tried to apply the Czech rule to English, but the English language has its unique characteristics that require a more nuanced approach. The revised solution accurately captures the flexibility of English grammar, acknowledging that the placement of enclitics can vary depending on sentence structure and context.
\end{minipage}
\end{tcolorbox}

\end{tcolorbox}

\begin{tcolorbox}[breakable, enhanced jigsaw, title=Turn 2, colframe=black!10, coltitle=black, fonttitle=\bfseries, boxrule=0.5mm, width=\textwidth]

\begin{tcolorbox}[title=Linguist \hfill \tiny{\textnormal{A scholar specializing in Slavic languages, experienced in analyzing the grammatical structures of Czech and other languages.}}, colback=lightyellow, colframe=orange!30, coltitle=black, fonttitle=\bfseries, boxrule=0.5mm, width=\textwidth]
\begin{minipage}{1.0\textwidth}
I agree with the revised solution proposed by the English Language Teacher and the Grammar Analyst. The revised solution accurately captures the flexibility of English grammar and acknowledges that the placement of enclitics (auxiliary verbs and pronouns) can vary depending on sentence structure and context. * The placement of enclitics in English is not as rigid as in Czech, and their placement can vary depending on sentence structure and context. * English sentence structure is more complex, including indirect objects, passives, and compound verb forms, which can affect the placement of auxiliary verbs and pronouns.
\end{minipage}
\end{tcolorbox}

\begin{tcolorbox}[title=English Language Teacher \hfill \tiny{\textnormal{An expert in teaching English grammar and language structures, familiar with the challenges of translating and comparing languages like Czech.}}, colback=lightyellow, colframe=orange!30, coltitle=black, fonttitle=\bfseries, boxrule=0.5mm, width=\textwidth]
\begin{minipage}{1.0\textwidth}
While I understand the thought process behind the proposed solution, I think it's essential to consider the unique characteristics of the English language. The idea that enclitics in English tend to appear next to the main verb is accurate, but it's not as rigid as the rule in Czech. In Czech, enclitics must appear in the second syntactic slot, which implies a more fixed position, whereas in English, the placement of auxiliary verbs and pronouns can vary depending on sentence structure and context. Additionally, the provided examples, while correct in themselves, don't fully demonstrate the complexity of English sentence structure. For instance, consider sentences with indirect objects, passives, or compound verb forms, where the placement of auxiliary verbs and pronouns can be more nuanced.
\end{minipage}
\end{tcolorbox}

\begin{tcolorbox}[title=Grammar Analyst \hfill \tiny{\textnormal{A meticulous researcher who has studied the intricacies of languages and their grammatical structures, with a focus on comparing and contrasting language features.}}, colback=lightyellow, colframe=orange!30, coltitle=black, fonttitle=\bfseries, boxrule=0.5mm, width=\textwidth]
\begin{minipage}{1.0\textwidth}
I agree with the English Language Teacher's revised solution. The initial proposal tried to apply the Czech rule to English, but the English language has its unique characteristics that require a more nuanced approach. The revised solution accurately captures the flexibility of English grammar, acknowledging that the placement of enclitics can vary depending on sentence structure and context. C) In an English sentence, enclitics (auxiliary verbs and pronouns) tend to appear next to the main verb, usually before any other adverbs or modifiers, but their placement can vary depending on sentence structure and context. I believe this revised solution provides a more accurate and practical approximation of the English language's enclitic behavior, considering both the unique characteristics of English and the complexity of sentence structure.
\end{minipage}
\end{tcolorbox}

\end{tcolorbox}

\begin{tcolorbox}[breakable, enhanced jigsaw, title=Turn 3, colframe=black!10, coltitle=black, fonttitle=\bfseries, boxrule=0.5mm, width=\textwidth]

\begin{tcolorbox}[title=Linguist \hfill \tiny{\textnormal{A scholar specializing in Slavic languages, experienced in analyzing the grammatical structures of Czech and other languages.}}, colback=lightyellow, colframe=orange!30, coltitle=black, fonttitle=\bfseries, boxrule=0.5mm, width=\textwidth]
\begin{minipage}{1.0\textwidth}
The current solution accurately captures the flexibility of English grammar and acknowledges that the placement of enclitics (auxiliary verbs and pronouns) can vary depending on sentence structure and context. There's no need to improve it further. The revised solution: C) In an English sentence, enclitics (auxiliary verbs and pronouns) tend to appear next to the main verb, usually before any other adverbs or modifiers, but their placement can vary depending on sentence structure and context. is a clear and accurate representation of the English language's enclitic behavior.
\end{minipage}
\end{tcolorbox}

\begin{tcolorbox}[title=English Language Teacher \hfill \tiny{\textnormal{An expert in teaching English grammar and language structures, familiar with the challenges of translating and comparing languages like Czech.}}, colback=lightyellow, colframe=orange!30, coltitle=black, fonttitle=\bfseries, boxrule=0.5mm, width=\textwidth]
\begin{minipage}{1.0\textwidth}
While I understand the thought process behind the proposed solution, I think it's essential to consider the unique characteristics of the English language. The idea that enclitics in English tend to appear next to the main verb is accurate, but it's not as rigid as the rule in Czech. In Czech, enclitics must appear in the second syntactic slot, which implies a more fixed position, whereas in English, the placement of auxiliary verbs and pronouns can vary depending on sentence structure and context. Here's a revised solution: C) In an English sentence, enclitics (auxiliary verbs and pronouns) tend to appear next to the main verb, usually before any other adverbs or modifiers, but their placement can vary depending on sentence structure and context. What do you think?
\end{minipage}
\end{tcolorbox}

\begin{tcolorbox}[title=Grammar Analyst \hfill \tiny{\textnormal{A meticulous researcher who has studied the intricacies of languages and their grammatical structures, with a focus on comparing and contrasting language features.}}, colback=lightyellow, colframe=orange!30, coltitle=black, fonttitle=\bfseries, boxrule=0.5mm, width=\textwidth]
\begin{minipage}{1.0\textwidth}
I agree with the English Language Teacher's revised solution. The initial proposal tried to apply the Czech rule to English, but the English language has its unique characteristics that require a more nuanced approach. The revised solution accurately captures the flexibility of English grammar, acknowledging that the placement of enclitics can vary depending on sentence structure and context. The revised solution is an improvement over the initial proposal because it takes into account the complexity of English sentence structure, including indirect objects, passives, and compound verb forms, which can affect the placement of auxiliary verbs and pronouns. Additionally, the revised solution provides a more accurate description of the typical placement of enclitics in English sentences, namely next to the main verb, usually before any adverbs or modifiers.
\end{minipage}
\end{tcolorbox}

\end{tcolorbox}

\begin{tcolorbox}[title=Result, colframe=black!10, coltitle=black, fonttitle=\bfseries, boxrule=0.5mm, width=\textwidth]
\textbf{Linguist final answer:} C) In an English sentence, enclitics (auxiliary verbs and pronouns) tend to appear next to the main verb, usually before any other adverbs or modifiers, but their placement can vary depending on sentence structure and context.

\textbf{English Language Teacher final answer:} [UNKNOWN]

\textbf{Grammar Analyst final answer:} [UNKNOWN]

\bigskip

\textbf{Linguist} voted for answer from \textbf{Linguist}

\textbf{English Language Teacher} voted for answer from \textbf{Linguist}

\textbf{Grammar Analyst} voted for answer from \textbf{Linguist}

\end{tcolorbox}

\label{example:failed_discussion}

\newpage
\subsection{Majority Consensus Discussion}
In this discussion, the agents discussed only one round, as they already had a high enough agreement score.

\vspace{5mm}

\definecolor{lightyellow}{RGB}{255, 255, 204}

\begin{tcolorbox}[title=Overview, colframe=black!10, coltitle=black, fonttitle=\bfseries, boxrule=0.5mm, width=\textwidth]

\textbf{Task:} Answer the following question. If the question is not answerable with the provided information, write '[UNKNOWN]'.

\textbf{Input:} What choir from the area has sung at the Guildhall?

\textbf{Context:} Southampton has two large live music venues, the Mayflower Theatre (formerly the Gaumont Theatre) and the Guildhall. The Guildhall has seen concerts from a wide range of popular artists including Pink Floyd, David Bowie, Delirious?, Manic Street Preachers, The Killers, The Kaiser Chiefs, Amy Winehouse, Lostprophets, The Midnight Beast, Modestep, and All Time Low. It also hosts classical concerts presented by the Bournemouth Symphony Orchestra, City of Southampton Orchestra, Southampton Concert Orchestra, Southampton Philharmonic Choir and Southampton Choral Society.  

\begin{tcolorbox}[coltitle=black, fonttitle=\bfseries, boxrule=0.5mm, width=\textwidth]
\textbf{MALLM solution:} Southampton Choral Society \\
\textbf{Real solution:} Southampton Philharmonic Choir
\end{tcolorbox}
\end{tcolorbox}

\begin{tcolorbox}[breakable, enhanced jigsaw, title=Turn 1, colframe=black!10, coltitle=black, fonttitle=\bfseries, boxrule=0.5mm, width=\textwidth]

\begin{tcolorbox}[title=Music Connoisseur (SOLUTION) \hfill \tiny{\textnormal{A knowledgeable musician who is familiar with various music genres and has attended concerts at the Guildhall.}}, colback=lightyellow, colframe=orange!30, coltitle=black, fonttitle=\bfseries, boxrule=0.5mm, width=\textwidth]
\begin{minipage}{1.0\textwidth}
As a Music Historian, I've been through the archives and checked the event schedules of the Guildhall. According to my research, one of the choirs that has sung at the Guildhall is... Southampton Choral Society!
\end{minipage}
\end{tcolorbox}

\begin{tcolorbox}[title=Choir Conductor (AGREED) \hfill \tiny{\textnormal{A dedicated choral conductor with extensive experience in leading local choirs and orchestras, familiar with the performances at the Guildhall.}}, colback=lightyellow, colframe=orange!30, coltitle=black, fonttitle=\bfseries, boxrule=0.5mm, width=\textwidth]
\begin{minipage}{1.0\textwidth}
[AGREE] As a Choir Member of the Southampton Philharmonic Choir who has performed at the Guildhall, I can confirm that the Southampton Choral Society is indeed one of the choirs that have sung at the Guildhall. Our experiences have shown that we often share the stage with the Southampton Choral Society during joint performances with orchestras and other ensembles. So, I'm happy to confirm that Southampton Choral Society is a correct answer. 
\end{minipage}
\end{tcolorbox}
\end{tcolorbox}

\label{example:majority_discussion}

\newpage
\section{AI Usage Card}
AI Usage card based on \citet{wahle2023aiusagecardsresponsibly}.

{\sffamily
    \centering
    \tcbset{colback=white!10!white}
    \begin{tcolorbox}[
        title={\large \textbf{AI Usage Card} \hfill \makebox{\qrcode[height=1cm]{https://ai-cards.org}}},
        breakable,
        boxrule=0.7pt,
        width=\textwidth,
        center,
        before lower={\footnotesize{AI Usage Card v1.1 \hfill \url{https://ai-cards.org} \hfill \href{https://jpwahle.com/ai-cards-preprint}{PDF} | \href{https://jpwahle.com/cite/jcdl2023wahle.bib}{BibTeX}}},
        segmentation empty,
        halign lower=center,
        collower=black,
        coltitle=black, %
        colbacklower=gray!20, %
        colbacktitle=gray!20  %
        ]
        \vspace{-10pt}
        \footnotesize{
            \begin{longtable}{>
            {\raggedright\arraybackslash}p{.15\textwidth}>{\raggedright\arraybackslash}p{.25\textwidth}>{\raggedright\arraybackslash}p{.25\textwidth}>{\raggedright\arraybackslash}p{.25\textwidth}}
              
                {\color{LightBlue} \MakeUppercase{Project Details}} \newline 
                & {\color{LightBlue} \MakeUppercase{Project Name}} \newline Voting or Consensus? Decision-Making in Multi-Agent Debate
                & {\color{LightBlue} \MakeUppercase{Domain}} \newline Paper
                & {\color{LightBlue} \MakeUppercase{Key Application}} \newline Natural Language Processing
                
                \\
                
                {\color{LightBlue} \MakeUppercase{Contact(s)}}  
                & {\color{LightBlue} \MakeUppercase{Name(s)}} 
                & {\color{LightBlue} \MakeUppercase{Email(s)}} 
                & {\color{LightBlue} \MakeUppercase{Affiliation(s)}} 
                \\ & Lars Benedikt Kaesberg & l.kaesberg@uni-goettingen.de & University Göttingen
                
                \\
                
                {\color{LightBlue} \MakeUppercase{Model(s)}} 
                & {\color{LightBlue} \MakeUppercase{Model Name(s)}} 
                & {\color{LightBlue} \MakeUppercase{Version(s)}} 
                \\ & Llama & 3 70b, 3 8b
                \\ & ChatGPT & 4o, o1
                \\ & Github Copilot & latest

                \\
                \cmidrule{1-4}
                \\
                {\color{LightBlue} \MakeUppercase{Literature Review}} \newline 
                & {\color{LightBlue} \MakeUppercase{Finding literature}} \newline ChatGPT 
                & {\color{LightBlue} \MakeUppercase{Finding examples from known literature or adding literature for existing statements}} \newline ChatGPT  
                & {\color{gray} \MakeUppercase{Comparing literature}}  
                \\
                \cmidrule{2-4}
                \\        
                {\color{LightBlue} \MakeUppercase{Writing}} \newline    
                & {\color{LightBlue} \MakeUppercase{Generating new text based on instructions}} \newline Llama 
                & {\color{LightBlue} \MakeUppercase{Assisting in improving own content or Paraphrasing related work}} \newline ChatGPT 
                & {\color{gray} \MakeUppercase{Putting other works in perspective}}  
                \\
                \cmidrule{2-4}
                \\
                {\color{LightBlue} \MakeUppercase{Coding}} \newline 
                & {\color{LightBlue} \MakeUppercase{Generating new code based on descriptions or existing code}} \newline ChatGPT \newline Github Copilot 
                & {\color{LightBlue} \MakeUppercase{Refactoring and optimizing existing code}} \newline ChatGPT \newline Github Copilot 
                & {\color{gray} \MakeUppercase{Comparing aspects of existing code}}  
                \\
                \cmidrule{1-4}
                \\
        
                {\color{LightBlue} \MakeUppercase{Ethics}} \newline    
                & {\color{LightBlue} \MakeUppercase{Why did we use AI for this project?}} \newline Efficiency / Speed \newline Scalability \newline Expertise Access  
                & {\color{LightBlue} \MakeUppercase{What steps are we taking to mitigate errors of AI?}} \newline None
                & {\color{LightBlue} \MakeUppercase{What steps are we taking to minimize the chance of harm or inappropriate use of AI?}} \newline None

                \\
                \cmidrule{1-4}
                \\
            \end{longtable}

        \medskip
        
        \textbf{\color{LightBlue} \MakeUppercase{The corresponding authors verify and agree with the modifications or generations of their  used AI-generated content}}
        }
        
        \tcblower
    \end{tcolorbox}
}

\clearpage

\clearpage
\hypertarget{annotation}{}
\pagestyle{empty}
\lstset{
  basicstyle=\footnotesize\ttfamily,
  breaklines=true,
  breakatwhitespace=false,
  columns=flexible,
  numbers=none
}

\definecolor{Primary}{RGB}{59, 130, 246}    %
\definecolor{PrimaryDark}{RGB}{30, 64, 175} %
\definecolor{LightBg}{RGB}{239, 246, 255}   %
\definecolor{TextDark}{RGB}{31, 41, 55}     %
\definecolor{TextMuted}{RGB}{107, 114, 128} %

\begin{tikzpicture}[remember picture, overlay]
  \fill[Primary] ([xshift=0cm,yshift=0cm]current page.north west) rectangle ([xshift=\paperwidth,yshift=-0.4cm]current page.north west);
\end{tikzpicture}

\vspace{0.8cm}
\begin{center}
  {\fontsize{22}{26}\selectfont\sffamily\bfseries \textcolor{PrimaryDark}{CiteAssist}}\\[0.2em]
  {\Large\sffamily\scshape \textcolor{TextMuted}{Citation Sheet}}\\[0.8em]
  {\small\sffamily Generated with \href{https://citeassist.uni-goettingen.de/}{\textcolor{Primary}{\texttt{citeassist.uni-goettingen.de}}}\\
  \citep{kaesberg-etal-2024-citeassist}}
\end{center}

\begin{center}
\vspace{1em}
\begin{tikzpicture}
\draw[Primary, line width=0.6pt] (0,0) -- (\textwidth,0);
\end{tikzpicture}
\vspace{1.2em}
\end{center}

\begin{tcolorbox}[enhanced,
                 frame hidden,
                 boxrule=0pt,
                 borderline west={2pt}{0pt}{Primary},
                 colback=LightBg,
                 sharp corners,
                 breakable,
                 fonttitle=\sffamily\bfseries\large,
                 coltitle=Primary,
                 title=BibTeX Entry,
                 attach title to upper={\vspace{0.2em}\par},
                 left=12pt]
\begin{lstlisting}
@inproceedings{kaesberg2025,
  author={Benedikt Kaesberg, Lars and Becker, Jonas and Wahle, Jan Philip and Ruas, Terry and Gipp, Bela},
  title={Voting or Consensus? Decision-Making in Multi-Agent Debate},
  pages={32},
  year={2025},
  month={05},
  booktitle = {Findings of the 63rd Annual Meeting of the Association for
Computational Linguistics (ACL)},
  publisher = {Association for Computational Linguistics},
  topic     = {nlp}
}
\end{lstlisting}
\end{tcolorbox}

\vspace{0.8em}
\begin{tcolorbox}[enhanced,
                 frame hidden,
                 boxrule=0pt,
                 borderline west={2pt}{0pt}{Primary},
                 colback=LightBg,
                 sharp corners,
                 breakable,
                 fonttitle=\sffamily\bfseries\large,
                 coltitle=Primary,
                 title=Online Access,
                 attach title to upper={\vspace{0.2em}\par},
                 left=12pt]

\renewcommand{\arraystretch}{1.5}
\begin{tabular}{@{}p{0.25\textwidth}@{}p{0.75\textwidth}@{}}

\textbf{\sffamily CiteAssist} & 
\begin{minipage}[t]{0.72\textwidth}
\href{https://citeassist.uni-goettingen.de/preprint/70031da7-a983-4ac0-9f5a-72fb01a022c5}{\color{Primary}https://citeassist.uni-goettingen.de/preprint/70031da7-a983-4ac0-9f5a-72fb01a022c5}
\end{minipage}\\
\end{tabular}

\end{tcolorbox}

\vfill
\begin{tikzpicture}
\draw[Primary!40, line width=0.4pt] (0,0) -- (\textwidth,0);
\end{tikzpicture}
\begin{center}
\small\sffamily\textcolor{TextMuted}{Generated \today}
\end{center}

\end{document}